\documentclass[pra,usletter,superscriptaddress,twocolumn]{revtex4}
\usepackage{amsmath,latexsym,amssymb,amsfonts}

\newcommand{\baseRing}[1]{\ensuremath{\mathbb{#1}}} 
 
\newcommand{\Sy}{\baseRing{S}} 
 
\newcommand{\R}{\baseRing{R}}

\newcommand{\ri}{\mathop{\mathrm{ri}}} 
 
\newcommand{\range}{\mathop{\mathrm{range}}}

\begin{document}

\preprint{} \title{A complete family of separability criteria}
\author{ Andrew C. Doherty} \affiliation{Institute for Quantum
  Information, California Institute of Technology, Pasadena, CA 91125,
  USA} \author{Pablo A. Parrilo} \affiliation{Automatic Control Laboratory, 
  ETH Zentrum, CH-8092 Z\"urich, Switzerland} 
\author{Federico M. Spedalieri}
\affiliation{Institute for Quantum Information, California Institute
  of Technology, Pasadena, CA 91125, USA}

\date{\today}

\begin{abstract}
  We introduce a new family of separability criteria that are based on
  the existence of extensions of a bipartite quantum state $\rho$ to a
  larger number of parties satisfying certain symmetry properties. It
  can be easily shown that all separable states have the required
  extensions, so the non-existence of such an extension for a
  particular state implies that the state is entangled. One of the
  main advantages of this approach is that searching for the extension
  can be cast as a convex optimization problem known as a semidefinite
  program (SDP).  Whenever an extension does not exist, the dual
  optimization constructs an explicit entanglement witness for the
  particular state. These separability tests can be ordered in a
  hierarchical structure whose first step corresponds to the
  well-known Positive Partial Transpose (Peres-Horodecki) criterion,
  and each test in the hierarchy is at least as powerful as the
  preceding one. This hierarchy is complete, in the sense that any
  entangled state is guaranteed to fail a test at some finite point in
  the hierarchy, thus showing it is entangled. The entanglement
  witnesses corresponding to each step of the hierarchy have
  well-defined and very interesting algebraic properties that in turn
  allow for a characterization of the interior of the set of positive
  maps. Coupled with some recent results on the computational
  complexity of the separability problem, which has been shown to be
  NP-hard, this hierarchy of tests gives a complete and also
  computationally and theoretically appealing characterization of
  mixed bipartite entangled states.
\end{abstract}
\pacs{03.67.Mn,03.65.Ud} \maketitle

\section{Introduction}

Entanglement is one of the most fascinating features of quantum
mechanics.  As Einstein, Podolsky and Rosen~\cite{einstein1935a}
pointed out, the quantum states of two physically separated systems
that interacted in the past can defy our intuitions about the outcomes
of local measurements. Entangled pure states have zero entropy but can
appear to have maximal entropy when the experimenter only has access
to one of the subsystems. On the other hand, Bell
inequalities~\cite{bell} quantify the extent to which local
measurements on separated quantum systems can be correlated in ways
that are forbidden in any local classical model. Violations of these
inequalities require entanglement. Moreover, it has recently been
recognized that entanglement is a very important resource in quantum
information processing, allowing certain important tasks such as
teleportation, quantum computation, quantum cryptography and quantum
communication to name a few~\cite{nielsen2000}.

For the case of pure states, determining when a given state is
entangled is very easy, since it is based on properties of the Schmidt
decomposition or, equivalently, the rank of the reduced density
matrices, which can be computed very efficiently. However, for the
case of mixed bipartite states, no single practical procedure that can
be guaranteed to detect the entanglement of every entangled state has
been found.  In the past few years, a considerable effort has been
dedicated to this
problem~\cite{lewenstein2000b,terhal2002a,bruss2002a,horodecki2001a}.
Still only incomplete criteria have been proposed, that can detect
some entangled states but not all of them or that work only for
certain restricted dimensions.  This is a somewhat uncomfortable
situation, since all the quantum states generated in the laboratory
for practical applications of quantum information processing are mixed
states. Hence the need, not only from the theoretical but also from
the practical point of view, of having an efficient tool that would
allow us to determine when a given state is entangled.

A bipartite mixed state is said to be separable~\cite{werner1989a}
(not entangled) if it can be written as a convex combination of pure
product states
\begin{equation}
\label{sep}
\rho =\sum p_{i}|\psi _{i}\rangle \langle \psi _{i}|\otimes |\phi
_{i}\rangle \langle \phi _{i}|,
\end{equation}
where $|\psi _{i}\rangle$ and $|\phi _{i}\rangle $ are state-vectors
on the spaces ${\mathcal{H}}_{A}$\ and ${\mathcal{H}}_{B}$ of
subsystems $A$ and $B$ respectively, and $p_i >0, \sum_i p_i =1$. If a
state admits such a decomposition, then it can be created by local
operations (unitary transformations, measurements, etc.) and classical
communication (LOCC) by the two parties, and hence it cannot be an
entangled state.  Despite the simplicity of (\ref{sep}), it has been
shown recently by Gurvits~\cite{gurvits2002a} that deciding whether or
not such a decomposition exists for a given density matrix is an
NP-hard problem.  This result destroys any hope of finding a
computationally efficient tool to determine entanglement of mixed
states as was the case for pure states, so long as the widely believed
result $P\neq NP$ is actually true.  But there are some instances of
the separability problem that allow efficient algorithms to solve
them. This is one of the basic ideas behind separability criteria.

A separability criterion is based on a simple property that can be
shown to hold for every separable state. They provide necessary but
not sufficient conditions for separability. If some state $\rho$ does
not satisfy the property, then it must be entangled. But if it does
satisfy it, that does not imply that the state is separable. One of
the first and most widely used of these criteria is the Positive
Partial Transpose (PPT) criterion, introduced by
Peres~\cite{peres1996a}.  If $\rho$ has matrix elements $\rho_{i k,j
  l} = \langle i|\otimes \langle k| \rho |j \rangle \otimes |l\rangle$
then the partial transpose $\rho^{T_A}$ is defined by
\begin{equation}
\rho^{T_A}_{i k,j l}=\rho_{j k,i l}.
\end{equation}
If a state is separable, then it must have a positive partial
transpose (PPT). To see this consider the decomposition (\ref{sep})
for $\rho$. Partial transposition takes $|\psi _{i}\rangle \langle
\psi _{i}|$ to $|\psi_{i}^*\rangle \langle \psi_{i}^*|$, so the
partial tranpose of $\rho$ can be written as
\begin{equation}
\label{ptsep}
\rho^{T_A} =\sum p_{i}|\psi_{i}^*\rangle \langle \psi_{i}^*|\otimes |\phi
_{i}\rangle \langle \phi _{i}|.
\end{equation}
Clearly $\rho^{T_A}$ is a valid quantum state and in particular it
must be positive semidefinite. Thus any state for which $\rho^{T_A}$
is not positive semidefinite is necessarily entangled. This criterion
is computationally very easy to check.  Furthermore, it was shown by
the Horodeckis~\cite{horodecki1996a}, based on previous work by
Woronowicz~\cite{woronowicz1976a}, to be both necessary and sufficient
for separability in ${\mathcal{H}}_{2} \otimes {\mathcal{H}}_{2}$ and
${\mathcal{H}}_{2}\otimes {\mathcal{H}}_{3}$. However, in higher
dimensions, there are PPT states that are nonetheless entangled, as
was first shown in~\cite{horodecki1997a}, again based
on~\cite{woronowicz1976a}.  These states are called bound entangled
states because they have the peculiar property that no entanglement
can be distilled from them by local operations~\cite{horodecki1998a}.

A different useful separability criterion, that has been used to show
entanglement of PPT states, is the range
criterion~\cite{woronowicz1976a,horodecki1997a}.  It is based on the
fact that for every separable state $\rho$ there exist a set of pure
product states $\{|\psi_i \rangle |\phi_i\rangle \}$ that span the
range of $\rho$ while $\{|\psi_i^* \rangle |\phi_i\rangle \}$ span the
range of $\rho^{T_A}$, as can be easily seen by looking at equations
(\ref{sep}) and (\ref{ptsep}). This criterion is sometimes stronger
than PPT, but in some cases it can be weaker (for example, when
considering full rank non-PPT states)..  Other criteria, that are in
general weaker than PPT are the reduction
criterion~\cite{horodecki1999b,cerf1999a} and the majorization
criterion~\cite{nielsen2001a}. None of these criteria, nor a
combination of them are sufficient to give a complete characterization
of separable states.

Another approach to distinguishing separable and entangled states
involves the so called {\it entanglement witnesses}
(EW)~\cite{terhal2000a}. An EW is an observable $W$ whose expectation
value is nonnegative on any separable state, but strictly negative on
an entangled state $\rho$.  We say in this case that $W$ ``witnesses''
the entanglement of $\rho$. Besides giving another theoretical tool to
detect entangled states, this idea addresses the question of whether
there is an experimental way of distinguishing an entangled state from
a separable one. By studying the geometrical structure of the set of
quantum states, it can be shown that for every entangled state there
exists an entanglement witness
$W$~\cite{woronowicz1976a,horodecki1996a}.  Thus, there is always an
observable that can be measured that will show that the state is
entangled.

There are two other important mathematical objects related to
entanglement witnesses. Although these do not have the physical
interpretation of observables they allow connections to other results
in the mathematical and mathematical physics literature. In the first
place, there is a correspondence that relates entanglement witnesses
to linear positive (but not completely positive) maps from operators
on $\mathcal{H}_A$ to operators on $\mathcal{H}_B$ (or vice versa);
see equation~(\ref{posmap}) and
reference~\cite{jamiolkowski1972a}. Applying such a map to one half of
an entangled state does not necessarily result in a positive
matrix. For this reason positive maps were rejected as possible
physical evolutions of quantum states in favor of the completely
positive maps. The PPT test has this structure where the transpose is
the positive map.  Any positive but not completely positive map
results in an analogous separability criterion. The equivalence
between entanglement witnesses and positive maps implies that if
$\rho$ is entangled there is always a positive map that will detect
the entanglement in this way~\cite{woronowicz1976a,horodecki1996a}.
The characterization of positive linear maps was in fact the original
motivation for studying the separability
question~\cite{woronowicz1976a}.

Finally, there is a well known mapping between positive linear maps
and positive semidefinite biquadratic
forms~\cite{choi1975a,jamiolkowski1974a}. This can be appreciated
simply by writing the condition that $W$ is positive on pure product
states explicitly in terms of the elements of $W$ and the state
vectors for the two systems, as in equation~(\ref{ewform}) in
Section~\ref{sec:cew}. This suggests the use of results from real
algebraic geometry~(see for example \cite{BCR98} and the references
therein) to attack the separability problem. Indeed, the semidefinite
programming techniques we employ here were first developed in this
general context~\cite{parrilo2003a}.

The question of whether a given state $\rho$ is separable may be
phrased as quantified polynomial inequalities in a finite
number of variables:
\begin{equation}
\label{entopt}
\forall W  [ 
\forall |\psi \rangle \forall |\phi \rangle \; 
\langle \psi |  \langle \phi | W |\psi \rangle |\phi
\rangle \geq 0 
\Longrightarrow
 {\mathrm{Tr}}[\rho W] \geq 0 
].
\end{equation}
If this proposition is satisfied then $\rho$ is separable. Since the
inequalities may be expressed in terms of polynomials of the variables
(the components of $W,|\psi \rangle, |\phi \rangle $) this is a
semi-algebraic problem. Much is known about the general class of
semi-algebraic problems, in particular the fact that they are
\emph{decidable}. The Tarski-Seidenberg decision procedure
\cite{BCR98} can then be used to provide an explicit algorithm to
solve the separability problem in all cases and therefore to decide
whether $\rho$ is entangled. A drawback of this approach is that most
exact techniques in algebraic geometry scale very poorly with the
number of variables (the Hilbert space dimensions in the separability
problem). For this problem, these methods do not perform well in
practice except for very small problem instances. This is in contrast
to the PPT test which may be implemented very efficiently but does not
always settle the question of separability of $\rho$. In this paper we
discuss a set of separability criteria that also have this property;
they all scale polynomially with the Hilbert space dimension and
perform well in practice, any state $\rho$ that is entangled is
detected by one of the tests but no one test detects all entangled
states.  Since the separability problem is NP-Hard it is very
unlikely that a procedure guaranteed to solve the problem in all
instances can scale well with Hilbert space dimension. As a result our
family of separability tests is, in some sense, the best way of
solving the problem from a practical point of view, in that simple
tests will detect the easiest instances of the problem, while the more
complicated instances genuinely require more computational resources.

The most important characteristic of the separability problem is the
fact that the separable states form a convex set. The existence of
entanglement witnesses, observables that are positive on separable
states but negative on some entangled state, is a direct result of
this convexity. There has been much work on the separability problem,
particularly from the Innsbruck-Hannover group as reviewed
in~\cite{terhal2002a,bruss2002a}, that emphasizes convexity and
proceeds by characterizing entanglement witnesses in terms of their
extreme points, the so-called optimal entanglement witnesses, and PPT
entangled states in terms of their extreme points, the edge PPT
entangled states~\cite{lewenstein2000c,lewenstein2001a}.  Convexity
also plays a central role in our work which provides a computational
means of constructing entanglement witnesses with certain properties.
It is interesting that our construction will allow us to characterize
the interior of the set of entanglement witnesses, but not its extreme
points. 

Beyond the separability problem, many problems of interest in quantum
information have the 
structure of convex optimizations~\cite{VaBbook}, a fact that
has found increasing
application in the field in recent years. One early example is the use
of results about linear programming to find the optimal local entanglement
concentration procedure for a pure bipartite state
in~\cite{jonathan1999a}. Our work will involve convex optimizations known as
semidefinite programs~\cite{VaBbook,VaB:96}, generalizations of linear
programs that optimize a linear function of a positive matrix
subject to linear constraints. Semidefinite programming arguments have
also been used been used in the quantum information
literature to address questions about quantum coin tossing,
distillation and optimal state
transformations~\cite{kitaevmsri,
  ambainis2003a,audenaert2002a,rains2001a,audenaert2001a}.

In this paper we discuss in detail a family of separability
criteria introduced in~\cite{doherty2002a}, that can be ordered into a
hierarchy of tests that have the following two very important
properties: i) the hierarchy is complete, i.e., any entangled state
will be detected by some test in the hierarchy, ii) there are
efficient computational algorithms to check each of the tests.  This
provides us with a very practical algorithmic way for testing
entanglement of a given bipartite mixed states, that is guaranteed to
detect any entangled state. Furthermore, the algorithm constructs an
explicit proof of this fact in the form of an entanglement witness.
This in turn helps us to develop a characterization of almost all
positive maps that are not completely positive.

The paper is organized as follows. In Section II we introduce a new
family of separability criteria; in Section III we introduce and
discuss the properties of semidefinite programs (SDP), and show that
each separability test in the family can be cast as a SDP, and briefly
discuss the resources needed to implement them.  In Section IV we
discuss how to take advantage of the symmetries that each test
requires to further reduce the computational resources needed. In
Section V we present an explicit proof of the completeness of the
hierarchy, translating previous results~\cite{fannes1988a,raggio1989a}
into the language of density matrices. Section VI shows how the
duality of the SDP can be exploited to construct an entanglement
witness that proves entanglement for a given state, and we discuss the
algebraic properties of these witnesses.  In Section VII we present
examples of the application of the hierarchy. In section VIII we
discuss how to construct an entangled state that is not detected by
the second test of the hierarchy and present several important
consequences of this result.  Section IX shows how to use an SDP to
test indecomposability of an entanglement witness and construct a
bound entangled state detected by it. In Section X we discuss the
connection betweeen entanglement witnesses and positive maps, and show
how the properties of the witnesses obtained through our hierarchy of
tests can be translated into a characterization of strictly positive
maps. Finally, in Section XI, we summarize our results and present our
conclusions.

\section{PPT Symmetric Extensions and new separability criteria}

Any separable state $\rho$ in ${\cal H}_A \otimes {\cal H}_B$ can be
written as in (\ref{sep}). Consider now the state $\tilde\rho$ in
${\cal H}_A \otimes {\cal H}_B \otimes {\cal H}_A$, given by
\begin{equation}
\label{extension}
\tilde\rho = \sum p_{i}|\psi _{i}\rangle \langle \psi _{i}|\otimes |\phi
_{i}\rangle \langle \phi _{i}|\otimes |\psi _{i}\rangle \langle \psi _{i}|.
\end{equation}
Then $\tilde\rho$ has the following properties: i) $\tilde\rho$ is an
extension of $\rho$ to three parties, in the sense that
\begin{equation}
\label{trace}
\mathrm{Tr_C} [\tilde\rho] = \rho,
\end{equation}
where $\mathrm{Tr_C}$ means that we take the partial trace over the
third party which we have taken to be equal to ${\cal H}_A$. ii)
$\tilde\rho$ is symmetric under interchanges of the first and third
parties, i.e., the two copies of party A. More precisely, if we define
the swap operator $P$ by
\begin{equation}
\label{swap}
P|i\rangle \otimes |k\rangle \otimes |j\rangle
=|j\rangle \otimes |k\rangle \otimes |i\rangle,
\end{equation}
the symmetry condition can be written as
\begin{equation}
\label{symmetry}
\tilde\rho = P \tilde\rho P.
\end{equation}
iii) $\tilde\rho$ must remain positive under any partial transposition
(since $\tilde\rho$ is also a separable state). Note that, due to the
symmetry (\ref{symmetry}), taking partial transpose with respect to
the third subsystem is equal to taking it with respect to the first
one. Now, for an {\it arbitrary} state $\rho$ in ${\cal H}_A \otimes
{\cal H}_B$, we will call $\tilde\rho$ a {\it PPT symmetric extension
  of $\rho$ to two copies of} ${\cal H}_A$ , if and only if,
$\tilde\rho$ satisfies the three properties stated above.  Since we
have shown by construction that any separable state has a PPT
symmetric extension to two copies of ${\cal H}_A$, then we can use its
existence as a separability criterion. If a given state does not have
such an extension, then the state must necessarily be entangled.

We can take this idea of the existence of PPT symmetric extensions
further by considering extending the state to an arbitrary number of
copies of subsystem A.  For any separable state in ${\cal H}_A \otimes
{\cal H}_B$ given by (\ref{sep}), the state
\begin{equation}
\label{extensionN}
\tilde\rho = \sum p_{i}|\psi _{i}\rangle \langle \psi _{i}|\otimes |\phi
_{i}\rangle \langle \phi _{i}|\otimes |\psi _{i}\rangle \langle \psi _{i}|^{\otimes n-1},
\end{equation}
is a state in ${\cal H}_A \otimes {\cal H}_B \otimes {\cal
  H}_A^{\otimes n-1}$ that, i) is symmetric under interchanges of any
two copies of subsystem A, ii) yields the original state $\rho$ in
${\cal H}_A \otimes {\cal H}_B$ when we trace out any $n-1$ copies of
subsystem A, and iii) remains positive under all possible partial
transpositions.  Again, for an arbitrary state $\rho$, we will call
$\tilde\rho$ a {\it PPT symmetric extension of $\rho$ to $n$ copies of
  party A}, if and only if, $\tilde\rho$ satisfies properties i), ii)
and iii). And as before, we can use the existence of this extension to
$n$ copies of subsystem A as a separability criterion. We have thus
generated a countably infinite family of separability criteria. Note
that the same idea can be generalized to the multipartite case: the
existence of PPT symmetric extensions to any number of copies of the
parties is a separability criterion.

For the bipartite case, these separability criteria are not completely
independent of each other, but they actually have a hierarchical
structure. We will now show that if a state has a PPT symmetric
extension to $n$ copies of A, call it $\tilde\rho_{n}$ , then it must
have a PPT symmetric extension to $n-1$ copies of A.  Let
$\tilde\rho_{n-1} = \text{Tr}_A [\tilde\rho_{n}]$, where $A$
represents one of the copies of $A$. It is easy to see that
$\tilde\rho_{n-1}$ will inherit from $\tilde\rho_{n}$ the property of
being symmetric under interchanges of copies of party A, since we have
just removed one of the copies. It is also obvious that
$\tilde\rho_{n-1}$ is an extension of $\rho$ to $n-1$ copies of A.
Let's assume that is not PPT. Then there is a subset $\mathcal{I}$ of
the parties such that $\tilde\rho_{n-1}^{T_{\mathcal{I}}}$ has a
negative eigenvalue, where $T_{\mathcal{I}}$ represents the partial
transpose with respect to all the parties in subset $\mathcal{I}$. Let
$|e\rangle$ be the corresponding eigenvector and let $\{|i\rangle\}$
be a basis of the system $A$ over which the partial trace was
performed. Since $\tilde\rho_{n}$ is PPT, then $\langle e|\langle
i|\tilde\rho_{n}^{T_{\mathcal{I}}}|e\rangle|i\rangle \geq 0$, for all
$i$. Then
\begin{equation}
\label{rhonpt}
\sum_i  \langle e|\langle i|\tilde\rho_{n}^{T_{\mathcal{I}}}|e\rangle|i\rangle =
\langle e|{\mathrm{Tr}}_A [\tilde\rho_{n}^{T_{\mathcal{I}}}]|e\rangle \geq 0.
\end{equation}
Since we performed the partial trace over a party that is not included
in $\mathcal{I}$ , we can commute the trace and the partial transpose,
and using $\tilde\rho_{n-1} = \text{Tr}_A [\tilde\rho_{n}]$, we have $
\langle e|\tilde\rho_{n-1}^{T_{\mathcal{I}}}|e\rangle \geq 0$, which
contradicts the fact that $|e\rangle$ is an eigenvector of
$\tilde\rho_{n-1}^{T_{\mathcal{I}}}$ with negative eigenvalue.

We have then constructed a family of separability criteria with a
natural hierarchical structure. If we take the usual PPT criterion as
the first step of the hierarchy, the existence of a PPT extension to
two copies of A as the second step, and so on, we see that the tests
are ordered in such a way that each test is at least as powerful as
the previous one, in the sense that if a state was shown to be
entangled by one of them, it will be also shown to be entangled by all
the tests that are higher in the hierarchy. This family of tests has
several very important and useful properties. It can be shown that
each test can be cast as a semidefinite program (SDP), which is a
class of convex optimization problems for which efficient algorithms
exist. The duality structure of the SDP allows us to construct an
explicit entanglement witness whenever a state fails one of the tests.
And finally, it can be proven that the hierarchy is {\it complete},
i.e., every entangled state is guaranteed to fail the test at some
finite point in the hierarchy.

\section{Semidefinite programs and searching for PPT symmetric extensions}

In this section we will introduce and discuss the structure of a
semidefinite program and we will show explicitly how to apply it to
the problem of searching for a PPT symmetric extension.

\subsection{Semidefinite programs}

A semidefinite program (SDP) is a particular type of convex
optimization problem~\cite{VaB:96,VaBbook}. An SDP corresponds to the
optimization of a linear function subject to a linear matrix
inequality (LMI).  A typical SDP has the form
\begin{eqnarray}
\label{sdp}
{\mathrm{minimize}} &\ \  c^T {\bf x}  \nonumber \\
{\mathrm{ subject \ to}} &\ \  F({\bf x}) \geq 0, 
\end{eqnarray}
where $c$ is a given vector, ${\bf x} = (x_1,\ldots,x_n)$, and $F({\bf
  x}) = F_0 + \sum_i x_i F_i$, for some fixed hermitian matrices
$F_i$. The inequality in the second line means that the matrix $F({\bf
  x})$ must be positive semidefinite. The minimization is performed
over the vector ${\bf x}$, whose components are the variables of the
problem. The set of feasible solutions, i.e., the set of $\mathbf{x}$
that satisfy the LMI, is a convex set. In the particular case in which
$c=0$, there is no function to minimize and the problem reduces to
whether or not the LMI can be satisfied for some value of the vector
${\bf x}$. In this case, the SDP is refered to as a {\it feasibility
  problem}. The convexity of the SDP has made it possible to develop
sophisticated and reliable analytical and numerical methods to solve
them~\cite{VaB:96}.

A very important property of a SDP, both from the theoretical and
applied points of view, is its {\it duality structure}. To any SDP of
the form (\ref{sdp}), which is usually called the {\it primal
  problem}, there is associated another SDP, called the {\it dual
  problem}, that can be stated as
\begin{eqnarray}
\label{dual}
{\mathrm{maximize}} &\ \ -\text{Tr} [F_0 Z]  \nonumber \\
{\mathrm{subject \ to}} &\ \  Z \geq 0 \nonumber \\
                       &\ \ \text{Tr} [F_i Z] = c_i,
\end{eqnarray}
where the matrix $Z$ is hermitian and is the variable over which the
maximization is performed. This corresponds to the maximization of a
linear functional, subject to linear constraints and a LMI.  Let ${\bf
  x}$ and $Z$ be any two feasible solutions of the primal and dual
problems respectively. Then we have the following relationship
\begin{equation}
\label{certif}
 c^T {\mathbf{x}}+ \text{Tr} [F_0 Z] = \text{Tr} [F({\bf x}) Z]  \geq 0,
\end{equation}
where the last inequality follows from from the fact that both $F({\bf
  x})$ and $Z$ are positive semidefinite. From (\ref{sdp}) and
(\ref{dual}) we can see that the left hand side of (\ref{certif}) is
just the difference between to objective functions of the primal and
dual problem. The inequality in (\ref{certif}) tells us that the value
of the primal objective function evaluated on any feasible vector
$\mathbf{x}$, is always greater or equal than the value of the dual
objective function evaluated on any feasible matrix $Z$. This property
is known as \emph{weak duality}. Thus, we can use any feasible
$\mathbf{x}$ to compute an upper bound for the optimum of $-\text{Tr}
[F_0 Z]$, and we can also use any feasible $Z$ to compute a lower
bound for the optimum of $c^T {\mathbf{x}}$.

If the feasibility constraints on both the primal and dual problems
are satisfied for some $Z>0$ and ${\bf x}$ such that $F({\bf x})>0$,
the problems are termed {\it strictly feasible}, and the optimum
values of the primal and dual formulations are equal. This property is
called \emph{strong duality}.  Furthermore, there is a feasible pair
$({\bf x}_{\rm opt},Z_{\rm opt})$ achieving the optimum. In this case,
as can be seen from equation (\ref{certif}), we have ${\rm Tr}\,
[F({\bf x}_{\rm opt})Z_{\rm opt}]=0$, and thus $F({\bf x}_{\rm
  opt})Z_{\rm opt}=0$, so the hermitian matrices $F({\bf x}_{\rm
  opt})$ and $Z_{\rm opt}$ have orthogonal ranges. This is known as
the {\em complementary slackness} condition~\cite{VaB:96}.

Equation (\ref{certif}) has another important application. Consider
the particular case of a feasibility problem (i.e., $c=0$). Then,
equation (\ref{certif}) will read
\begin{equation}
\label{feascert}
\text{Tr} [F_0 Z]  \geq 0,
\end{equation}
and this must hold for {\it any feasible solution} of the dual
problem.  This property can be used to give a {\it certificate of
  infeasibility} for the primal problem: {\textit{if there exists $Z$
    such that $Z\geq 0$ and ${\mathrm{Tr}} [F_i Z] = 0$, that
    satisfies ${\mathrm{Tr}} [F_0 Z] < 0$, then the primal problem
    must be infeasible.}} We will show later that for the particular
case of our hierarchy of separability tests, whenever a PPT symmetric
extension of $\rho$ cannot be found (primal problem is infeasible),
the certificate provided by the dual problem is nothing but an
entanglement witness for the state $\rho$.

\subsection{Separability tests as semidefinite programs}

Each test in the hierarchy of separability criteria introduced in
Section II can be written as a semidefinite program. We will show in
detail how the SDP is setup for the second test in the hierarchy,
which corresponds to searching for PPT symmetric extensions of $\rho$
to two copies of subsystem A. The general case, of extensions to $n$
copies of party A, can be constructed in a similar way.

Let $\{ \sigma_i^A \}_{i=1}^{d_A^2}, \{\sigma_j^B\}_{j=1}^{d_B^2}$ be
bases for the space of hermitian matrices that operate on ${\cal H}_A$
and ${\cal H}_B$, of dimensions $d_A$ and $d_B$ respectively, such
that they satisfy
\begin{equation}
\label{tracesig}
\text{Tr} [ \sigma_i^X \sigma_j^X] = \alpha_X \delta_{i j} \ \ \ \text{and}
\ \ \ \text{Tr} [\sigma_i^X]  =  \delta_{i1},
\end{equation}
where $X$ stands for $A$ or $B$, and $\alpha_X$ is some constant---the
generators of SU(n) could be used to form such a basis.  Then we can
expand $\rho$ in the basis $\{ \sigma_i^A \otimes \sigma_j^B \}$, and
write $\rho = \sum_{i j} \rho_{i j} \sigma_i^A \otimes \sigma_j^B$,
with $\rho_{i j}=\alpha_A^{-1}\alpha_B^{-1} {\mathrm{Tr}} [\rho \,
\sigma_i^A \otimes \sigma_j^B]$. In the same way, we can expand the
extension $\tilde\rho$ in ${\cal H}_A \otimes {\cal H}_B \otimes {\cal
  H}_A$ as
\begin{eqnarray}
\label{rhoext}
\tilde\rho & = & \sum_{\stackrel{i j k}{\tiny{i<k}}}
\tilde\rho_{ijk} \{\sigma_i^A 
\otimes \sigma_j^B \otimes \sigma_k^A +\sigma_k^A 
\otimes \sigma_j^B \otimes \sigma_i^A \} + \nonumber \\
& & + \sum_{kj}\tilde\rho_{kjk} \,\sigma_k^A 
\otimes \sigma_j^B \otimes \sigma_k^A, 
\end{eqnarray}
where we made explicit use of the swapping symmetry between the first
and third parties, that we require from $\tilde\rho$. To satisfy the
condition that $\tilde\rho$ is an extension of $\rho$, we need to
impose
\begin{equation}
\label{extcond}
\text{Tr}_C [\tilde\rho] = \rho,
\end{equation}
where $\text{Tr}_C$ means tracing out the third party. Using
(\ref{tracesig}) and the fact that $\{ \sigma_i^A \otimes \sigma_j^B
\}$ form a basis of ${\cal H}_A \otimes {\cal H}_B$, equation
(\ref{extcond}) reduces to
\begin{equation}
\label{fixvar}
\tilde\rho_{ij1} = \rho_{ij}.
\end{equation}
This fixes some of the components of $\tilde\rho$. The remaining ones
will play the role of the variables in the SDP. The LMIs come from
requiring that the extension $\tilde\rho$ and all its partial
tranposes are positive semidefinite. If we define
\begin{eqnarray}
\label{Fs}
G_0 & = & \sum_{j} \rho_{1j}\,\sigma_1^A \otimes\sigma_j^B \otimes\sigma_1^A 
+ \nonumber
\\
&&+ \sum_{i=2,j=1} \rho_{ij}\,\{\sigma_i^A \otimes\sigma_j^B \otimes\sigma_1^A +
\sigma_1^A \otimes\sigma_j^B \otimes\sigma_i^A\} \nonumber \\
G_{iji} & = &  \sigma_i^A \otimes\sigma_j^B \otimes\sigma_i^A 
\ \ \ \ \ \ \ \ \ \ \ \ \ \ \  i\geq 2,\nonumber \\
G_{ijk} & = &  (\sigma_i^A \otimes\sigma_j^B \otimes\sigma_k^A+
\sigma_k^A \otimes\sigma_j^B \otimes\sigma_i^A) 
 \ \ \ k> i\geq 2,\nonumber \\
&&
\end{eqnarray}
we can write the PSD condition $\tilde\rho \geq 0$ as
\begin{equation}
\label{LMI}
 G({\mathbf x}) = G_0 + \sum_J x_J G_J \geq 0,
\end{equation}
where we have collected all the subindices in (\ref{Fs}) into one
subindex $J$. Equation (\ref{LMI}) has exactly the form that appears
in (\ref{sdp}). The role of the variable $\bf{x}$ is played by the
coefficients $\tilde\rho_{ijk} (k \neq 1, k\geq i)$, which can vary
freely without affecting the extension condition (\ref{extcond}). The
number of free variables is $m=(d_A^4d_B^2-d_A^2d_B^2)/2$, and each
matrix $G_J$ has dimension $n= d_A^2 d_B$. Since $\tilde\rho$ is
symmetric under swaps of the first and third parties, there are only
two independent partial transpositions that can be applied to it,
which we can take as partial transposes with respect to the first and
second parties (one of the copies of A, and subsystem B). The
requirement that these two partial transposes are positive leads to
two more LMIs, given by
\begin{equation}
\label{morelmi}
\tilde\rho^{T_A} \geq 0 \ \ \ \mathrm{and}  \ \ \ \tilde\rho^{T_B} \geq 0,
\end{equation}
where the $G_J$ matrices for these two inequalities are related to the
matrices given in (\ref{Fs}) by the appropriate partial transposes,
namely $G_J^{T_A}$ and $G_J^{T_B}$.  We can actually combine the three
LMIs into one, by defining the matrix
\begin{equation}
\label{G}
F = \tilde\rho \oplus \tilde\rho^{T_A} \oplus \tilde\rho^{T_B},
\end{equation}
and using the fact that a block-diagonal matrix $C = A \oplus B$ is
positive semidefinite, if and only if, both $A$ and $B$ are positive
semidefinite.

We have then stated the search of a PPT symmetric extension of $\rho$
as an SDP, in which the objective function is zero ($c=0$), so it
corresponds to a feasibility problem, and the LMI condition reads $F =
\tilde\rho \oplus \tilde\rho^{T_A} \oplus \tilde\rho^{T_B} \geq 0$,
which encodes the requirement of the extension and its partial
transposes being positive semidefinite. The SDP will then take the
form
\begin{eqnarray}
\label{sdptest}
{\mathrm{minimize}} &\ \  0  \nonumber \\
{\mathrm{subject \ to}} &\ \  F  \geq 0. 
\end{eqnarray}  
In appendix A, we discuss a slightly modified version of the SDP that
has the advantage of performing better numerically, but we will keep
the form (\ref{sdptest}) for all the analytical discussions, since its
dual form is more clearly related to the construction of entanglement
witnesses, which is another one of the main results of this paper.

The SDP for other tests of the hierarchy (extensions to more copies of
party A), can be constructed in the same way, by generating the
matrices $G_J$ with the appropriate symmetry, and constructing the
block-diagonal matrix $F$, whose blocks correspond to all the
independent partial transposes that can be applied to the extension
$\tilde\rho$.

\subsection{Resources needed to implement the tests}

As we mentioned before, there are very efficient algorithms to solve
semidefinite programs, and we can use their properties to discuss the
computational resources required to implement a general step in our
hierarchy of tests. Assume that we are searching for a PPT symmetric
extension of a state $\rho$ in ${\cal H}_A \otimes {\cal H}_B$ to $k$
copies of subsystem A, with $d_A$ and $d_B$ the dimensions of ${\cal
  H}_A$ and ${\cal H}_B$ respectively. Then, the corresponding
semidefinite program will have
$m=\left[\left(\begin{smallmatrix}d_A^2+k-1 \\k
    \end{smallmatrix}
  \right)-d_A^2 \right] d_B^2 $ variables and a matrix $G$ with $
\left( k+1\right)$ blocks of dimension $d_A^{2 k} d_B^2$. Numerical
SDP solvers are described in detail in~\cite{VaB:96}. Typically they
involve the solution of a series of least squares problems each
requiring a number of operations scaling with problem size as
$O(m^2n^2)$, where $F(\mathbf{x})$ is an $n \times n$ matrix. For SDPs
with a block structure these break into independent parts each with a
value of $n$ determined by the block size. The number of iterations
required is known to scale no worse than $O(n^{1/2})$. Thus for any
fixed value of $k$ the computation involved in checking our criteria
scales no worse than $O(d_A^{13k/2})$ which is polynomial in the
system size. On the other hand, for $d_A$ and $d_B$ fixed, the size of
the matrix $F(\mathbf{x})$ scales exponentially with the number of
copies $k$. There is, however, a significant improvement that can be
accomplished by exploiting the swapping symmetry to its fullest. In
the next section we will show that we can impose a stronger
restriction on the extension that brings the scaling of resources down
to polynomial in the number of copies of subsystem A, for fixed $d_A$,
$d_B$.  It is important to point out that the resources required to
solve the separability problem have been proven to scale super
polynomially only when the dimensions of \emph{both} systems are
allowed to vary. These two results are consistent, although the
complexity result implies that there is no value of $k$ such that the
$k^{th}$ test detects all entangled states for all values of $d_A$,
$d_B$.

\section{Exploiting the symmetry}

Each test for separability searches for an extension of a state $\rho$
in ${\cal H}_A \otimes {\cal H}_B$ of dimension $d_A d_B$, to the
space ${\cal H}_A^{\otimes k} \otimes {\cal H}_B$, that has dimension
$d_A^k d_B$ (where, without loss of generality, we have interchanged
the order of ${\cal H}_B$ and all copies of A for convenience). We see
that the dimension of the extended space increases exponentially with
the number of copies of party A. We have shown that we can impose
further restrictions on the extension, and in particular we require it
to be invariant under swaps of the copies of subsystem A. This reduces
the size of the space over which we search for the extension, but the
scaling with the number of copies remains exponential, which is not
desirable of a practical tool for deciding separability of a state.
However, we can actually impose a stronger constraint on the form of
the extension, that reduces the scaling of its size from exponential
to polynomial in the number of copies.

As we pointed out before, any separable state in ${\cal H}_A \otimes
{\cal H}_B$ of the form (\ref{sep}) has a PPT symmetric extension to
${\cal H}_A^{\otimes k} \otimes {\cal H}_B$, that we can explicitly
write as
\begin{equation}
\label{sepext2}
\tilde\rho = \sum p_{i}|\psi _{i}\rangle \langle \psi _{i}|^{\otimes k}
\otimes |\phi_{i}\rangle \langle \phi _{i}|.
\end{equation}
This extension is obviously invariant under swaps of copies of $A$,
and we used this property to restrict the form of the matrices $F_J$
in the LMI of our SDP. But $\tilde\rho$ has a more constraining
property: {\it its support and range are contained in the symmetric
  subspace of ${\cal H}_A^{\otimes k}\otimes{\cal H}_B$} (where the
symmetry is understood to apply only to the copies of $A$). For the
case of the extension to two copies of system $A$, we can write the
projector into this symmetric subspace as $\pi = \frac{1}{2}(\openone
+ P)$, with $P$ the swap operator defined in (\ref{swap}). Then, the
symmetry requirement on the extension takes the form $\tilde\rho = \pi
\tilde\rho \pi$.

For an arbitrary $\rho$, we can now restrict our search to extensions
that satisfy this property. If $\{ S_i^A\}$ is a basis of hermitian
matrices having support and range in the symmetric subspace of ${\cal
  H}_A^{\otimes k}$, this restriction is equivalent to only
considering matrices $G$ in (\ref{LMI}) of the form $G= S_i^A\otimes
\sigma_j^B$.  Since the dimension of the symmetric subspace in ${\cal
  H}_A^{\otimes k}$ is
\begin{equation}
\label{dimsym}
d_{S_k} = \left(\begin{matrix}d_A+k-1 \\ k \end{matrix}
  \right),
\end{equation}
with $d_A$ the dimension of ${\cal H}_A$, the number of matrices of
this form is $d_{S_k}^2 d_B^2$. The number of variables in the SDP is
this number minus the number of constraints given by equation
(\ref{extcond}), which is $m=(d_{S_k}^2 - d_A^2) d_B^2$. By using
(\ref{dimsym}), we get $m=O(k^{(d_A -1)})$, which for a fixed size of
$A$ is only polynomial in the number of copies. Since the matrices
$G_J$ have range and support only on this symmetric subspace, we know
that by a suitable change of basis they can be simultaneously
block-diagonalized, with the only nonzero block having size $n =
d_{S_k}^2 d_B^2$.

The SDP that searches for the PPT extension also requires to check
positivity of a certain number of partial transposes of the extension
$\tilde\rho$. These checks translate into a bigger LMI, although we
will now show that this does not change the scaling properties of its
size. Consider the case in which we apply the partial transpose to the
first $l$ copies of $A$, which we will denote
${\tilde\rho}^{T_{A^{\otimes l}}}$. Since the matrices $G_J$ have
support and range only on the symmetric subspace of ${\cal
  H}_A^{\otimes k}$, it is not difficult to show that the matrices
$G_J^{T_{A^{\otimes l}}}$ must have support {\it only on the tensor
  product of a subspace isomorphic to the symmetric subspace of ${\cal
    H}_A^{\otimes l}$ and the symmetric subspace of ${\cal
    H}_A^{\otimes (k-l)}$.} The dimension of this tensor product is
just the product of the dimensions of the two subspaces.  If we
perform a change of coordinates by rotating to a basis that contains a
basis of this tensor product of symetric subspaces, we can see that
the size of the matrices $G^{T_{A^{\otimes l}}}$ can be taken as
$d_{S_l}^2 d_{S_{(k-l)}}^2$. This scales at most as $O(k^{2(d_A
  -1)})$. Since the number of independent partial tranposes is
$(k+1)$, as a result of the symmetry requirements, the size $n$ of the
matrices in the LMI scales not worse than $O(k^{2d_A -1})$.  Combining
this with the scaling of the number of variables $m$ shown above and
the scaling properties of solving the SDP, which is given by $O(m^2
n^2)$, we can see that for fixed $d_A$, the tests in the hierarchy
scale as $O(k^{(6 d_A -4)})$, which is polynomial on the number of
copies of party $A$ for a fixed $d_A$.

\section{Completeness of the hierarchy of tests}

One of the main results of this paper is the {\it completeness} of the
hierarchy of separability tests. This result allows us to give an
algorithm that will show if a state is entangled in a finite number of
steps (although this number may be high for some states). Even though
the hierarchy of tests is a new result, the proof of its completeness
is identical to the proof of certain properties of the possible
equilibrium states of a system that interacts with a thermal bath.
These results, which were proved by Raggio \textit{et
  al.}~\cite{raggio1989a}, and Fannes \textit{et
  al.}~\cite{fannes1988a}, have been in the literature for quite some
time.

It was noted in~\cite{werner1989b} that this
result~\cite{raggio1989a,fannes1988a} could be interpreted as a
characterization of bipartite quantum states, that requires that the
only states that can have symmetric extensions to any number of copies
of one of its subsystyems, are the {\it separable states.} The same
idea was independently conjectured recently by
Schumacher~\cite{schumacherpriv}.

We will present a proof of the completeness of the hierarchy, which is
basically the proof found in~\cite{fannes1988a}, applied to the case
of bipartite mixed states on finite dimensional spaces. Our discussion
has the same level of mathematical rigor and is based on the
techniques presented in the discussion of the Quantum de Finetti
Theorem in~\cite{caves2002a}. The theorem we will prove is stronger
than our hierarchy, since it requires the existence of symmetric
extensions without any requirements on the partial transposes. The
completeness of our hierarchy can be deduced from this result as a
corollary.

\newtheorem{thm1}{Theorem}
\begin{thm1}[Completeness]
  Let $\rho$ be a bipartite mixed state in ${\cal H}_A \otimes {\cal
    H}_B$.  Then $\rho$ has a symmetric extension to $k$ copies of
  subsystem A for any $k$, if and only if, $\rho$ is separable.
\end{thm1}
\textbf{Proof:} One of the implications is trivial. Assume $\rho$ is
separable.  Then we can write
\begin{equation}
\label{septhm}
\rho =\sum p_{i}|\psi _{i}\rangle \langle \psi _{i}|\otimes |\phi
_{i}\rangle \langle \phi _{i}|.
\end{equation}
>From this expression, we can write down explicitly a symmetric
extension $\tilde\rho$ for any value of $k$, namely
\begin{equation}
\label{extensionk}
\tilde\rho = \sum p_{i}|\psi _{i}\rangle \langle \psi _{i}|^{\otimes k-1}\otimes |\phi
_{i}\rangle \langle \phi _{i}|,
\end{equation}
and this completes the first part of the proof.

To prove the other implication, the idea is to use the existence of
the extensions to construct a set of states in ${\cal H}_A^{\otimes
  n}$ that can be shown to be separable by using the Quantum de
Finetti Theorem, and then show that this result implies that the
extensions themselves have to be separable.  Let $\rho$ be a state in
${\cal H}_A \otimes {\cal H}_B$ such that for any $n$, there is a
symmetric extension of $\rho$ in ${\cal H}_A^{\otimes n} \otimes {\cal
  H}_B$, which we will call $\tilde\rho_n$. Let us pick a fixed value
$k$ for the number of copies of party A. Let the set $\{ b_i
\}_{i=1}^{d_B^2}$ be a basis for the set of hermitian operators in
${\cal H}_B$, such that $b_i > 0$ for all $i$ (i.e., all these
operators are positive definite~\cite{posbasis}), and in particular
let us choose $b_1 = {\openone}_B$, the identity in ${\cal H}_B$.  Now
we define the operator
\begin{equation}
\label{reduct}
{\bar{\rho}}_{b_i ,k} =  \text{Tr}_B \left[({\openone}_A^{\otimes k} 
\otimes b_i) \tilde\rho_k 
\right],
\end{equation}
where ${\openone}_A$ is the identity on subsystem A.  The operator
${\bar{\rho}}_{b_i ,k}$ is positive semidefinite (PSD) and non zero
since all the operators $b_i$ were taken to be strictly positive. Then
${\bar{\rho}}_{b_i ,k}$ is proportional to a state in ${\cal
  H}_A^{\otimes k}$, since it is hermitian and PSD. We can choose the
operators $b_i$ such that $\text{Tr} [{\bar{\rho}}_{b_i ,k}] =1$ for
all $k$, so that (\ref{reduct}) is actually a normalized state in
${\cal H}_A^{\otimes k}$.

We will now prove that the existence of symmetric extensions
$\tilde\rho_k$ of $\rho$ for all $k$, imply that we can choose the
states ${\bar{\rho}}_{b_i ,k}$ to be {\it
  exchangeable}~\cite{caves2002a}.  Recalling the definition of
exchangeability we need to show that, for any $l > 0$, there are
states ${\bar{\rho}}_{b_i ,(k+l)}$ that are symmetric and satisfy
\begin{equation}
\label{marginal}
{\bar{\rho}}_{b_i ,k} =  \mathrm{Tr}_{A_{k+1} \ldots A_{k+l}} [ {\bar{\rho}}_{b_i ,(k+l)}].
\end{equation}
Let us fix $k$ and assume that there is not an extension
$\tilde\rho_k$ such that the state ${\bar{\rho}}_{b_i ,k}$ given by
(\ref{reduct}) is exchangeable. That means that there has to be a
value $l_1$ for which equation (\ref{marginal}) is not satisfied for
any ${\bar{\rho}}_{b_i ,k}$ and ${\bar{\rho}}_{b_i ,(k+l_1)}$. But
since $\rho$ has symmetric extensions for all $k$, we can just choose
an extension to $(k+l_1)$ copies $\tilde\rho_{(k+l_1)}$, and we have
that $\mathrm{Tr}_{A_{k+1} \ldots A_{k+l_1}} [\tilde\rho_{(k+l_1)}]$
is a symmetric extension of $\rho$ to $k$ copies, and
\begin{eqnarray}
{\bar{\rho}}_{b_i ,k} & = & \mathrm{Tr}_B \left[({\openone}_A^{\otimes k} 
\otimes b_i) \mathrm{Tr}_{A_{k+1} \ldots A_{k+l_1}}
[\tilde\rho_{(k+l_1)}]\right] \nonumber \\
& = & \mathrm{Tr}_{A_{k+1} \ldots A_{k+l_1}} \left[\mathrm{Tr}_B [({\openone}_A^{\otimes 
k+l_1} \otimes b_i) \tilde\rho_{(k+l_1)}]\right] \nonumber \\
& = &  \mathrm{Tr}_{A_{k+1} \ldots A_{k+l_1}} \left[{\bar{\rho}}_{b_i ,(k+l_1)}\right].
\end{eqnarray}
This is a contradiction, so we can conclude that we can always choose
the states ${\bar{\rho}}_{b_i ,k}$ to be exchangeable.

The state ${\bar{\rho}}_{b_i ,k}$ satisfies then the hypothesis of the
Quantum de Finetti Theorem~\cite{caves2002a}, and so we know there is
a \emph{unique} probability measure function $P_{b_i} (\varrho) \geq
0$, such that
\begin{equation}
\label{definetti}
{\bar{\rho}}_{b_i ,k} = \int_D \varrho^{\otimes k} P_{b_i} (\varrho) d\varrho,
\end{equation}
where $D$ represents the space of states in ${\cal H}_A$ (i.e., the
set of hermitian, positive semidefinite operators of trace one).

For each $\varrho$, we can think of $P_{b_i}(\varrho)$ as a functional
applied to the operators $b_i$, which we will note $F_{\varrho}$,
defined as $F_{\varrho} (b_i) = P_{b_i}(\varrho)$. This functional is
{\it linear} on convex combinations of positive operators.  To see
this, let $\mu > 0$. Then $F_{\varrho}(\mu b_i + (1-\mu) b_j) = P_{\mu
  b_i + (1-\mu) b_j}(\varrho)$, where $P_{\mu b_i + (1-\mu) b_j}$ is
the unique probability density that satisfies
\begin{eqnarray}
\label{linear2}
{\bar{\rho}}_{(\mu b_i + (1-\mu) b_j),k} & = & \int_D \varrho^{\otimes k} 
P_{\mu b_i + (1-\mu) b_j}(\varrho) d\varrho \nonumber \\
 & = & \text{Tr}_B \left[ ({\openone}_A^{\otimes k} \otimes ({\mu b_i + (1-\mu) b_j})) 
\tilde\rho_k \right]
\nonumber \\
 & = & \mu \text{Tr}_B \left[ ( {\openone}_A^{\otimes k}\otimes b_i) \tilde\rho_k \right]+
\nonumber \\ 
& & + (1-\mu) \text{Tr}_B \left[ ({\openone}_A^{\otimes k}\otimes b_j) \tilde\rho_k \right] 
\nonumber \\
 & = & \int_D \left(\mu  P_{b_i}(\varrho) + (1-\mu) P_{b_j}(\varrho)\right) 
\varrho^{\otimes k} d\varrho.\nonumber \\
&&
\end{eqnarray}
The second equality in (\ref{linear2}) holds because we are
considering a convex combination of the operators $b_i$, which
guarantees that $\text{Tr}_B [ ({\openone}_A^{\otimes k}\otimes ({\mu
  b_i + (1-\mu) b_j})) \tilde\rho_k ]$ is normalized.  Then, by the
uniqueness of the probability density in the Quantum de Finetti
Theorem, we have
\begin{equation}
\label{linear4}
P_{\mu b_i + (1-\mu) b_j} (\varrho) = \mu P_{b_i}(\varrho) + (1-\mu) P_{b_j}(\varrho),
\end{equation}
which translates into
\begin{equation}
\label{linear5}
F_{\varrho}({\mu b_i + (1-\mu) b_j}) = \mu F_{\varrho}(b_i) + (1-\mu)
 F_{\varrho}(b_j).
\end{equation}
Then $F_{\varrho}$ is a linear functional on convex combinations of
positive states in ${\cal H}_B$.

Since $F_{\varrho}$ is defined on a basis, there is a unique way of
extending this functional linearly to the whole space of operators in
${\cal H}_B$. So we have a linear, positive and continuous functional
on a finite dimensional Hilbert space, and it is a well-known result
that any such functional can be written as
\begin{equation}
\label{innerprod}
F_{\varrho}(b) = \text{Tr}_B [ {\bar{\sigma}}_{\varrho} b ] \ \ \ \forall b,
\end{equation}
for some {\it unique} positive semidefinite operator
${\bar{\sigma}}_{\varrho}$ in ${\cal H}_B$. This operator might not be
a state in ${\cal H}_B$ since it need not be normalized. We can then
define a function
\begin{equation}
\label{Prho}
P(\varrho) = {\mathrm{Tr}}[{\bar{\sigma}}_\varrho],
\end{equation}
that is nonnegative. If $P(\varrho)$ is nonzero, we can define
$\sigma_\varrho = {\bar{\sigma}}_{\varrho} / P(\varrho)$.  Then
(\ref{innerprod}) takes the form
\begin{equation}
\label{inprod}
P_b (\varrho) = F_{\varrho}(b) = \mathrm{Tr}_B [ {\sigma}_{\varrho} b ] P(\varrho) 
\ \ \ \forall b.
\end{equation}
Note that since $\sigma_{\varrho}$ is normalized, $P(\varrho) =
P_{\openone_B} (\varrho)$, which shows that $P(\varrho)$ is a
probability density.  Using (\ref{inprod}) in (\ref{definetti}), we
get
\begin{eqnarray}
\label{decomp1}
{\bar{\rho}}_{b_i ,k} & = & \int_D \varrho^{\otimes k} \,\mathrm{Tr}_B [ 
\sigma_{\varrho} b_i ]\,
 P (\varrho) d\varrho \nonumber \\
 & = & \mathrm{Tr}_B \left[(\openone_A^{\otimes k} \otimes b_i ) \int_D 
\varrho^{\otimes k} \otimes
\sigma_{\varrho}\,  P (\varrho) d\varrho \right].
\end{eqnarray}
If $P(\varrho)=0$ for some $\varrho$, we can define $\sigma_\varrho$
arbitrarily, since it would not contribute to the integral in
(\ref{decomp1}).  Since (\ref{decomp1}) is valid for all the elements
$b_i$ of a basis of hermitian matrices in ${\cal H}_B$, by comparing
the expression in the second line with (\ref{reduct}), we can deduce
that
\begin{equation}
\label{decomp2}
\tilde\rho_k = \int_D \varrho^{\otimes k} \otimes
\sigma_{\varrho}\,  P (\varrho) d\varrho.
\end{equation}
This means that $\tilde\rho_k$ is a separable state, since
(\ref{decomp2}) is an explicit decomposition as a convex combination
of product states.  Furthermore, since $\tilde\rho_k$ is an extension
of our original state $\rho$, we have
\begin{eqnarray}
\label{rhoissep}
\rho & = & \text{Tr}_{A_2 \ldots A_k} [\tilde\rho_k] \nonumber \\
     & = &  \int_D \varrho \otimes \sigma_{\varrho} \,  P (\varrho) 
d\varrho,
\end{eqnarray}
which shows that $\rho$ has to be a separable state. This concludes
the proof of the theorem. $\Box$

It is clear that this theorem implies the completeness of the
hierarchy of separability tests introduced in Section II, since a
state that has PPT symmetric extensions to $k$ copies of party $A$ for
all values of $k$ obviously has {\it symmetric extensions for all
  values of $k$}, which according to the theorem implies that the
state must be separable.  However, it is interesting to note that the
PPT requirement is not essential for the completeness of the
hierarchy. Searching just for symmetric extensions is also a complete
family of separability criteria and one that requires less resources.

In~\cite{terhal2002b} local hidden variable (LHV) theories were also
constructed for quantum states possessing so-called symmetric
quasi-extensions, where rather than requiring that the extension be
positive as a matrix it is only required that it is positive on
product states. The number of extensions corresponds to the number of
independent local measurement settings that the theory is able to
describe. In fact our argument that only separable states have an
arbitrary number of symmetric extensions generalizes to this
case.  Essentially all that is needed is a version
of the Quantum de Finetti theorem that holds for entanglement
witnesses as well as states but it is straightforward to check that
the argument of~\cite{caves2002a} holds in this case also since only
POVMs that act as tensor products on each subsystem are used in the
proof. Hence although the use of quasi-extensions is strictly stronger
for a small number of local measurement settings, if the LHV is
required to work for an arbitrary number of local measurement
settings, the construction will only work for separable states.

We could generate more families of criteria, by searching for
symmetric extensions that have to satisfy some other constraint, and
this family of tests would still be complete although it would in
general require more resources. If the constraint can be written in
terms of linear equalities and LMIs, we could still use an SDP to
implement the tests. Choosing between these many possibilities is a
matter of how well they perform in actual examples. It becomes a
trade-off between how much more powerful the tests become when more
constraints are placed on the extensions, and how much this increases
the resources needed. Including the PPT requirement on the extension
has the advantage that it guarantees that the second and higher tests
in the hierarchy are stronger than the PPT criterion, and we have
found this to be a good trade-off in practice.

\section{Construction of entanglement witnesses}
\label{sec:cew}

An entanglement witness (EW) for a state $\rho$ is a hermitian
operator $W$ that satisfies
\begin{equation}
\label{ew}
\mathrm{Tr} [\rho W] < 0 \ \ \ \text{and} \ \ \ \mathrm{Tr} [ \rho_{sep} W] \geq 0,
\end{equation}
where $\rho_{sep}$ is any separable
state~\cite{terhal2000a,horodecki1996a}.  It is clear that if
(\ref{ew}) is satisfied, then $\rho$ cannot be separable, and $W$
gives a proof of that fact. This property has a very nice geometric
interpretation. Since the set of separable states is convex, any point
that does not belong to it (like any entangled state), can be
separated from the set by an hyperplane. In our case, the operator $W$
defines the hyperplane. This result is known as the Hahn-Banach
theorem~\cite{rockafellar1970a}. In practice, finding a $W$ satisfying
$\text{Tr} [\rho W] < 0$ is not difficult, but proving $\mathrm{Tr} [
\rho_{sep} W] \geq 0$ might be very hard. To understand the reason for
this, let us recall that any separable state can be written as a
convex combination of projectors into pure product states
\begin{equation}
\label{sepxy}
\rho_{sep}=\sum_i p_{i}|x\rangle \langle x|\otimes |y\rangle \langle y|,
\end{equation}
where $|x\rangle = \sum_i x_i |i\rangle$, $|y\rangle = \sum_j y_j
|j\rangle$, for some bases $\{|i\rangle\}$ and $\{|j\rangle\}$ of
${\cal H}_A$ and ${\cal H}_B$, respectively. Then, $\mathrm{Tr}
[\rho_{sep} W]\geq 0$ for any separable state $\rho_{sep}$, if and
only if, $\mathrm{Tr} [|x\rangle \langle x|\otimes |y\rangle \langle
y| W]\geq 0$ for any product state $|x\rangle \langle x|\otimes
|y\rangle \langle y|$. Then we have
\begin{eqnarray}
\label{ewform}
E_W(x,y) = \langle x y |W |x y \rangle & = & \mathrm{Tr} [|x\rangle \langle x|\otimes 
|y\rangle \langle y|W]\nonumber \\
 & = &  \sum_{ijkl} W_{ijkl} x^*_i y^*_j x_k y_l.
\end{eqnarray}
We can the interpret then the requirement that $\mathrm{Tr}
[\rho_{sep} W] \geq 0$ as a positivity condition on the bihermitian
form $E_W(x,y)$ associated with the entanglement witness $W$, where
bihermitian means that the form is hermitian with respect to $x$ and
hermitian with respect to $y$.  It is a well-known result that
checking positivity of an arbitrary real form is an NP-hard problem,
and the result in~\cite{gurvits2002a} implies the same is true for
bihermitian forms. This is the reason why constructing entanglement
witnesses is not easy in general.

As we mentioned in Section III, any primal SDP has an associated dual
problem that is also a SDP, and in particular, whenever the primal
problem is infeasible, the dual problem provides a certificate of this
infeasibility.  We will show that in the case of our separability
tests, this infeasibility certificate generated by the dual problem is
actually an entanglement witness.

Consider the SDP (\ref{sdptest}), and let us focus on the second test
of the hierarchy, i.e., searching for PPT symmetric extensions to two
copies of party A. In this case, the dual problem takes the form
\begin{eqnarray}
\label{dual2}
{\mathrm{maximize}} &\ \ -\text{Tr} [F_0 Z]  \nonumber \\
{\mathrm{subject \ to}} &\ \  Z \geq 0, \nonumber \\
                       &\ \ \text{Tr} [F_i Z] = 0,
\end{eqnarray}
where $F_0$ has three blocks that encode the extension and its two
independent partial transposes, and from (\ref{G}) we can see that it
has the form
\begin{equation}
\label{G0}
F_0 = G_0 \oplus G_0^{T_A} \oplus G_0^{T_B}.
\end{equation}
Due to this block structure, we can restrict the search over $Z$ in
the dual program, to $Z$ that have the same structure, so we can take
\begin{equation}
\label{Zstruct}
Z = Z_0 \oplus Z_1^{T_A} \oplus Z_2^{T_B},
\end{equation}
where the $Z_i$ are operators in ${\cal H}_A \otimes {\cal H}_B
\otimes {\cal H}_A$. The positivity condition on $Z$ in (\ref{dual2}),
translates into a positivity requirements for each of the blocks in
(\ref{Zstruct}).  Using this structure we can write
\begin{equation}
\label{tracef0}
\text{Tr} [F_0 Z] = \text{Tr} [G_0 (Z_0 +Z_1 +Z_2)],
\end{equation}
since $\text{Tr} [G_0^{T_X} Z_i^{T_X}] = \text{Tr} [G_0 Z_i]$, for
$i=1,2$ and $X= A, B$. We defined $G_0$ in equation (\ref{Fs}) as a
linear function of $\rho$, so we can write $G_0 = \Lambda (\rho)$,
where $\Lambda$ is a linear map from operators on ${\cal H}_A \otimes
{\cal H}_B$ to operators on ${\cal H}_A \otimes {\cal H}_B \otimes
{\cal H}_A$, whose action on an arbitrary operator $Y$ on ${\cal H}_A
\otimes {\cal H}_B$ is given by
\begin{eqnarray}
\label{lambda}
\Lambda \left( Y\right) & =& Y\otimes \openone_{A}/d_{A}+P
\left( Y\otimes  \openone_{A}\right)
P/d_{A} \nonumber \\
& & \ \ \ \ \ \ \ -\openone_{A}\otimes \mathrm{Tr}_{A}\left[ Y\right] \otimes 
\openone_{A}
/d_{A}^{2},
\end{eqnarray}
where $P$ is the swap operator defined by $P|i\rangle \otimes
|k\rangle \otimes |j\rangle =|j\rangle \otimes |k\rangle \otimes
|i\rangle $.  We can now define an operator $\tilde{Z}$ on ${\cal H}_A
\otimes {\cal H}_B$ given by $\tilde{Z} = \Lambda^\ast (Z_0 + Z_1 +
Z_2)$, where $\Lambda^{\ast}$ is the adjoint map of $\Lambda$ and is
defined as the map that satisfies $\mathrm{Tr}[\Lambda(X)
Y]=\mathrm{Tr}[X \Lambda^\ast(Y)]$ for any hermitian operators $X ,
Y$. For our particular case, this map takes the form
\begin{eqnarray}
\label{lambdaadj}
\Lambda ^{\ast }\left( V\right) & = & \mathrm{Tr}_{C}\left[ V\right] /d_{A}+\mathrm{Tr
}_{C}\left[ P V P\right] /d_{A} \nonumber \\
&& \ \ \ \ \ \ \ -\openone_{A}\otimes \mathrm{Tr}_{AC}\left[ V\right]
/d_{A}^{2}.
\end{eqnarray}
Then we have
\begin{equation}
\label{trrhoZ}
\mathrm{Tr} [\rho \tilde{Z}] = \mathrm{Tr}[\Lambda (\rho)
(Z_0 + Z_1 + Z_2)] = \mathrm{Tr} [F_0 Z].
\end{equation}

Let $\rho_{sep}$ be any separable state. Then we know that there is a
PPT symmetric extension of $\rho_{sep}$, or equivalently, the primal
problem (\ref{sdptest}) is feasible. Then from (\ref{certif}), and
using the fact that $c = 0$, we have that $\mathrm{Tr} [F_0 Z] \geq 0$
for all dual feasible $Z$ and so from (\ref{trrhoZ}) we have
\begin{equation}
\label{trrhosep}
\mathrm{Tr} [\rho_{sep} \tilde{Z}] \geq 0,
\end{equation}
{\it for any $\tilde{Z}$ obtained from a feasible dual solution $Z$.}
This means that any operator $\tilde{Z}$ constructed in this way,
satisfies one of the two properties required in (\ref{ew}), and is
therefore a candidate for an entanglement witness.

Now consider the case in which the primal problem is not feasible for
a given state $\rho$. This can only occur if this state is entangled.
We can then use the arguments presented in appendix \ref{apb} to
affirm that there must be a feasible dual solution $Z_{EW}$ that
satisfies $\mathrm{Tr} [F_0 Z_{EW}] < 0$.  Using (\ref{trrhoZ}) and
(\ref{trrhosep}) we can see that the corresponding hermitian operator
${\tilde{Z}}_{EW}$ satisfies the two conditions
\begin{equation}
\label{Zew}
\mathrm{Tr} [\rho {\tilde{Z}}_{EW}] < 0 \ \ \ \mathrm{and} \ \ \ 
\mathrm{Tr} [\rho_{sep} {\tilde{Z}}_{EW}] \geq 0,
\end{equation}
which means that ${\tilde{Z}}_{EW}$ is an entanglement witness for the
state $\rho$.

Even though we have shown the calculation explicitly only for the
second test of the hierarchy, similar reasoning can be applied to all
tests to show that if the primal problem is infeasible, there is a
dual feasible solution that can be used to construct an entanglement
witness for the state $\rho$. The EWs obtained for each of the tests
have very well-defined and interesting algebraic properties, that can
also be used to interpret each step in the hierarchy as a search for
EWs of a particular form.

\subsection{Algebraic properties of the entanglement witnesses}

For any EW there is an associated bihermitian form given by
(\ref{ewform}). We have shown that the requirement that an
entanglement witness W is positive on all separable states, is
equivalent to requiring the associated form $E(x,y)$ to be positive.

Let us consider an EW obtained form the first test in the hierarchy,
which corresponds to the usual PPT criterion. It is a well-known
result that all states that fail this criterion, can be shown to be
entangled by an entanglement witness of the form
\begin{equation}
\label{decomposable}
W = P + Q^{T_A},
\end{equation}
where both $P$ and $Q$ are positive semidefinite operators.
Entanglement witnesses that have this form are called {\it
  decomposable}. If we note by $|\psi_p\rangle$ the eigenvectors of
$P$ and by $|\phi_p\rangle$ the eigenvectors of $Q$, we can write
\begin{eqnarray*}
P & = & \sum_p \kappa_p |\psi_p\rangle\langle \psi_p|, \nonumber \\
Q & = & \sum_p \lambda_p |\phi_p\rangle\langle \phi_p|,
\end{eqnarray*}
where the eigenvalues $\kappa_p$ and $\lambda_p$ are nonnegative,
since both $P$ and $Q$ are PSD.  If we study the associated form $E_W
(x,y)$, we have
\begin{widetext}
\begin{eqnarray}
\label{decompform}
E_W (x,y) & = & \langle x y |(P + Q^{T_A}) |x y \rangle =  \sum_p | 
\sqrt{\kappa_p} \langle \psi_p | x y\rangle|^2 + 
\sum_p |\sqrt{\lambda_p} \langle \phi_p | x^* y\rangle|^2  \nonumber \\
 & = & \sum_p | \sqrt{\kappa_p} \sum_{ij} \psi^p_{ij} x_i y_j|^2 + 
\sum_p |\sqrt{\lambda_p}  \sum_{ij} \phi^p_{ij} x_i^* y_j|^2,  
\end{eqnarray}
\end{widetext}
with $|\psi_p\rangle = \sum_{ij} \psi^p_{ij} |ij\rangle$ and
$|\phi_p\rangle = \sum_{ij} \phi^p_{ij} |ij\rangle$.  The last
equality in (\ref{decompform}) shows that $E_W(x,y)$ can be written as
a {\it sum of squared magnitudes} (SOS), which proves its positivity.
This property is an alternative description of decomposable
entanglement witnesses.

Now imagine that we have a state $\rho$ that is PPT entangled, whose
entanglement is detected by the second test of the hierarchy (i.e.,
$\rho$ does not have a PPT symmetric extension to two copies of party
A).  Then we know that the dual SDP will provide us with an
entanglement witness ${\tilde{Z}}_{EW}$ for this state. Let us
concentrate on the properties of this ${\tilde{Z}}_{EW}$. First, it is
clear that it cannot be decomposable, since decomposable EWs can only
detect states that are not PPT. By setting $\rho_{sep} = |x y \rangle
\langle x y|$ in (\ref{trrhosep}), we have that
\begin{eqnarray}
\label{Zewform}
\mathrm{Tr} [|x y \rangle \langle x y| {\tilde{Z}}_{EW}] & = & \langle x y |
{\tilde{Z}}_{EW}| x y \rangle \nonumber \\
 & = & E_{{\tilde{Z}}_{EW}} (x,y) \geq 0.
\end{eqnarray}
According to equation (\ref{trrhoZ}), we have
\begin{equation}
\label{trxyZew}
\mathrm{Tr} [|x y \rangle \langle x y| {\tilde{Z}}_{EW}] = \mathrm{Tr}[\Lambda 
(|x y \rangle \langle x y|) (Z_0 + Z_1 + Z_2)].
\end{equation}
The operator $\Lambda$ maps a state $\rho$ in ${\cal H}_A \otimes
{\cal H}_B$ into an operator in ${\cal H}_A \otimes {\cal H}_B \otimes
{\cal H}_A$ that is invariant under swaps of the two copies of A and
yields the original state $\rho$ when one of the copies of A is traced
out, but is in general not positive semidefinite.  Now consider the
state $|x y x\rangle \langle x y x|$. This state is invariant under
swaps of copies of system A and also satisfies $\text{Tr}_C [|x y
x\rangle \langle x y x|] = |x y \rangle \langle x y|$. Then we know
that there must exist some coefficients $a_J$ such that
\begin{equation}
\label{xyxexp}
|x y x\rangle  \langle x y x| = \Lambda (|x y \rangle \langle x y|)
+\sum_J a_J G_J,
\end{equation}
since the $G_J$ form a basis of the space of matrices $M$ satisfying
the swapping symmetry and $\text{Tr}_C [M] = 0$.  According to
(\ref{dual2}) we have $\text{Tr} [G_J Z_i] =0$, and hence we can
rewrite (\ref{trxyZew}) as
\begin{equation}
\label{trxyZew2}
\mathrm{Tr} [|x y \rangle \langle x y| {\tilde{Z}}_{EW}] = \mathrm{Tr}[|x y x \rangle 
\langle x y x| (Z_0 + Z_1 + Z_2)].
\end{equation}
Combining (\ref{Zewform}) and (\ref{trxyZew2}), we have
\begin{equation}
\label{forms}
\langle x y |{\tilde{Z}}_{EW}| x y \rangle = \langle x y x|(Z_0 + Z_1 + Z_2)
|x y x\rangle.
\end{equation}
Since we are working with normalized states, we know that $\langle
x|x\rangle =1$, so we can multiply the left-hand side of (\ref{forms})
by this factor without changing the equality, obtaining
\begin{equation}
\label{forms2}
E_{{\tilde{Z}}_{EW}} (x,y) \langle x|x\rangle =  \langle x y x|(Z_0 + Z_1 + Z_2)
|x y x\rangle.
\end{equation}
This equation is, in principle, only valid when the variables $x_i$
and $y_i$ correspond to a normalized state, i.e., when $\sum_i |x_i|^2
=1$ and $\sum_i |y_i|^2 =1$. However, since both sides of
(\ref{forms2}) are homogeneous functions of fourth degree on the
$x_i$, and of second degree on the $y_i$, we can extend this equality
to all values of the variables, and interpret (\ref{forms2}) as an
equality between two forms {\it that is satisfied everywhere}.  But we
can now rewrite the right-hand side of (\ref{forms2}) as
\begin{widetext}
\begin{equation}
\label{sos1}
\langle x y x|(Z_0 + Z_1 + Z_2)|x y x\rangle = \langle x y x|Z_0| x y 
x\rangle +\langle x^* y x|Z_1^{T_A}| x^* y 
x\rangle +\langle x y^* x|Z_2^{T_B}| x y^* 
x\rangle.
\end{equation}
\end{widetext}
Since $Z_0,Z_1^{T_A}$ and $Z_2^{T_B}$ are positive by construction,
equation (\ref{sos1}) gives an explicit sum of squares (SOS)
decomposition of the right-hand side of (\ref{forms2}). We can
conclude then that even though the form $E_{{\tilde{Z}}_{EW}} (x,y)$
is not a SOS, {\it it becomes a SOS when multiplied by the strictly
  positive SOS form $\langle x|x\rangle = \sum_i |x_i|^2$}.  This
property holds for {\it any} EW obtained from the second test.

This result generalizes to all steps of the hierarchy: {\it the
  bihermitian form associated with an EW obtained from the
  $(k+1)^{th}$ test of the hierarchy, can be written as a SOS when
  multiplied by the SOS form $\langle x|x\rangle^k = (\sum_i
  |x_i|^2)^k$}. We will say then that these EW are $k$-$SOS$. Then for
example, an entanglement witness that is $0$-$SOS$ is decomposable,
since its associated form can be written as a SOS (we will use SOS
instead of $0$-$SOS$ for this particular case). It is clear that if an
EW is $k$-$SOS$, it is also $l$-$SOS$ for all $l\geq k$. Note that for
$k\geq 1$, all $k$-$SOS$ entanglement witnesses are
\emph{indecomposable}.

As we discussed in Section V, searching for symmetric extensions with
no PPT requirement generates another complete family of separability
criteria. For this family, equation (\ref{forms2}) takes the form
\begin{equation}
\label{sos2}
E_{{\tilde{Z}}_{EW}} (x,y) \langle x|x\rangle =  \langle x y x|Z_0|x y x\rangle,
\end{equation}
since now the LMI has only one block, corresponding to the positivity
requirement on the extension. Since $Z_0$ is PSD, the right-hand side
of (\ref{forms3}) still is a SOS, so we will still say that
${\tilde{Z}}_{EW}$ is $\mathit{1}$-$SOS$ (or $k$-$SOS$ if we replace
$\langle x|x\rangle$ by $\langle x|x\rangle^k$ in (\ref{sos2})).  The
main difference between (\ref{sos2}) and (\ref{sos1}) is in the type
of terms that appear in the sum of squares decomposition. Note that
(\ref{sos2}) involves only squares of polynomials in the variables
$(x_i,y_j)$ while the second and third terms in the right-hand side of
(\ref{sos1}) correspond to squares of polynomials in the $(x_i^*, y_j
, x_k)$ and $(x_i,y_j^*)$ variables respectively. This situation
extends to all the steps of the hierarchy. The SOS decomposition
generated by the PPT family involves the squared magnitudes of all
possible polynomials in the variables $(x_i,y_j)$ {\it and their
  conjugates} that are compatible with the symmetry requirements,
while the SOS decomposition obtained from the non-PPT family involves
squared magnitudes of polynomials involving {\it only} the variables
$(x_i,y_j)$.  The completeness theorem tells us that for any EW
obtained from any of the two families, we could make the associated
bihermitian form into a SOS by multiplying by a certain power of the
SOS form $(\sum_i |x_i|^2)$.  However, the value of the power needed
in the non-PPT case will be in general higher than for the PPT case.

We presented our family of separability criteria as the search for an
extension of the original state $\rho$ that satisfied certain
symmetries and had positive partial transposes, and showed that this
search could be put in the primal form of a SDP. From the discussion
above, we can see that by looking at the dual SDP, we can interpret
this hierarchy as a search over possible entanglement witnesses for
the state $\rho$ that are $k$-$SOS$ for some $k$. Note that this
interpretation also applies to the non-PPT family of criteria, the
only change being the type of terms allowed in the SOS decomposition.
The completeness result proved in Section V tells us that the set of
all entanglement witnesses that are $k$-$SOS$ for some $k$, is {\it
  sufficient} for proving entanglement of any state. This raises a
very interesting question: \emph{are all EW} $k$-$SOS$ \emph{for some}
$k$? We will now show that most EW are, and the only ones that may not
be are those that are extremal in the sense that their associated
hyperplane touches the set of separable states. These are the {\it
  optimal} EWs from~\cite{bruss2002a}.

Let us first introduce some definitions regarding convex sets.  A set
$K$ is said to be a {\it convex cone} if it is convex and closed under
linear combinations with nonnegative coefficients, i.e., if $x , y \in
K$ and $ a , b \geq 0$, then $a x + b y \in K$. The {\it dual cone} of
$K$ is defined as $K^* = \{ z\, :\, \langle z , x\rangle \geq 0,
\forall x\in K\}$, where $\langle , \rangle$ represents some inner
product (note that $K^*$ will be different for different inner
products).  It is easy to show that $K^*$ is actually a closed convex
cone even if $K$ is not closed. An important property is
that~\cite{rockafellar1970a}
\begin{equation}
\label{dualdual}
\left( K^* \right)^* = \mathrm{cl} (K),
\end{equation}
where $\mathrm{cl} (K)$ represents the closure of $K$.

Let $S$ be the set of all unnormalized separable states. $S$ is a
closed convex cone. Its dual cone is $S^* = \{
Z\,:\,\mathrm{Tr}[Z\rho_{sep}] \geq 0, \forall \rho_{sep} \in S\}$,
which contains the set of all entanglement witnesses. If $(S^*)^o$
notes the {\it interior} of $S^*$, we have $(S^*)^o = \{ Z\, : \,
\mathrm{Tr}[Z\rho_{sep}] > 0, \forall \rho_{sep} \in S\}$.
\newtheorem{thm2}[thm1]{Theorem}
\begin{thm2}
  Let $W$ be an entanglement witness such that $W\in (S^*)^o$. Then
  $W$ is $k$-$SOS$ for some $k$, i.e., $\exists k$ such that $E_W
  (x,y) (\sum_i |x_i|^2 )^k \ {\mathrm{is \ a}} \, SOS$.
\end{thm2}
\textbf{Proof:} Let $O_k = \{ Z\,:\,E_Z(x,y)(\sum_i |x_i|^2 )^k \ 
{\mathrm{is \ a}}\, SOS\}$.  This is just the set of entanglement
witnesses that are $k$-$SOS$.  Clearly, $O_k \subset O_{k+1}$ and $O_k
\subset S^*$. Now we define the set
\begin{equation}
\label{defS}
O = \bigcup_{k=0}^\infty O_k.
\end{equation}
$O$ is a convex cone, although it may not be closed. We will now show
that the dual of this cone is the set $S$. Let $\rho_{sep} \in S$. For
any $Z \in O$, $\exists k$ such that $Z \in O_k$. But $Z \in S^*$, so
$\mathrm{Tr}[Z \rho_{{sep}}]\geq 0$, which means that $\rho_{{sep}}
\in O^*$, so we have
\begin{equation}
\label{incl1}
S \subset O^*.
\end{equation}
Now, let $\rho \in O^*$ and assume $\rho \notin S$; then $\rho$ is an
entangled state. By the completeness of the hierarchy of separability
tests, we know that there is a value of $k$ for which $\mathrm{Tr}[Z
\rho] <0$ for some $Z \in O_k \subset O$, and then we must have $\rho
\notin O^*$, which is a contradiction. Then
\begin{equation}
\label{incl2}
O^* \subset S.
\end{equation}
>From (\ref{incl1}) and (\ref{incl2}), we have $S = O^*$.  Then we can
use (\ref{dualdual}) to state that $S^* = \mathrm{cl}(O)$, which means
\begin{equation}
\label{intS*}
\left(S^*\right)^o \subset  O.
\end{equation}
If $W \in (S^*)^o$ then by (\ref{defS}) there exists $k$ such that $W
\in O_k$, and hence it is $k$-$SOS$. $\Box$

This theorem has a very nice geometric interpretation. It says that
the sequence of convex cones $O_k$ approximates the convex cone of all
entanglement witnesses $S^*$ from the inside, giving a complete
characterization of its interior in terms the $k$-$SOS$ property. On
the other hand, the entanglement witnesses on the boundary of $S^*$
may not be $k$-$SOS$ for any $k$. They satisfy $ \mathrm{Tr}[Z
\rho_{sep}] = 0$ for some separable state $\rho_{sep}$, and correspond
to the {\it optimal} entanglement witnesses discussed
in~\cite{bruss2002a}.

As we briefly mentioned in Section II, the separability criteria based
on searching for certain extensions of a state can easily be
generalized to the study of multipartite entanglement. The dual
formulation of searching for an EW with certain algebraic properties
also clearly applies to the multipartite case. It is known that
multipartite entanglement cannot be characterized in terms of
bipartite entanglement alone~\cite{bennett1999a, dur2000a}. However,
our approach can be generalized to the multipartite case in order to
construct another sequence of tests that is also \emph{complete}.
These results will be reported elsewhere~\cite{doherty2003c}.

\section{Examples}

We present now some examples for which we applied our techniques to
prove entanglement of certain PPT entangled states, and to construct
the appropriate entanglement witnesses. For all these examples, the
second test of the hierarchy (searching for PPT symmetric extensions
to two copies of party A) was sufficient to show entanglement.  We
used MATLAB to code the corresponding SDP, and used the package
SEDUMI~\cite{sturm} to solve it. The code is available at
\texttt{http://www.iqi.caltech.edu/publications.html}.

\subsection{$3 \otimes 3$ state.}
\label{choiex}

We consider the following state,described in~\cite{horodecki2001a},
given by
\begin{equation}
\label{choiform}
\rho_\alpha = \frac{2}{7} | \psi_+ \rangle \langle \psi_+ | +
\frac{\alpha}{7} \sigma_+ +
\frac{5-\alpha}{7} V \sigma_+ V,   
\end{equation}
with $0 \leq \alpha \leq 5$, $|\psi_+\rangle = \frac{1}{\sqrt{3}}
\sum_{i=0}^2 |i i \rangle$, $\sigma_+ = \frac{1}{3}(|0 1 \rangle
\langle 0 1 | + |1 2 \rangle \langle 1 2 | + |2 0 \rangle \langle 2 0
|)$ and $V$ the operator that swaps the two systems (note they are
both the same space).  Notice that $\rho_\alpha$ is invariant under
the simultaneous change of $\alpha \rightarrow 5-\alpha$ and
interchange of the parties. The state is separable for $2 \leq \alpha
\leq 3$ and not PPT for $\alpha > 4$ and $\alpha < 1$, which was
proved in~\cite{horodecki2001a} by using a positive map that is not
completely positive due to Choi~\cite{choi1975a}.  Our code solves the
SDP for this state in about 5 seconds on a desktop computer. From this
solution, numerical entanglement witnesses can be constructed for
$\rho_{\alpha}$ in the range $3+\epsilon<\alpha\leq 4$ (and
$1\leq\alpha<2-\epsilon$) with $\epsilon\geq 10^{-8}$. A witness for
$\alpha>3$ can be extracted from these by inspection:
\begin{widetext}
\begin{equation}
\label{choiwit}
\tilde{Z}_{EW} = 
2 \, (|0 0 \rangle \langle 0 0 | +
   |1 1 \rangle \langle 1 1 | +
   |2 2 \rangle \langle 2 2 |) + |0 2 \rangle \langle 0 2 | + 
|1 0 \rangle \langle 1 0 | + 
|2 1 \rangle \langle 2 1 | -
3 | \psi_+ \rangle \langle \psi_+ |.
\end{equation}
>From this entanglement witness, the Choi form~\cite{choi1975a} can be
recovered.  This observable is nonnegative on separable states:
\begin{eqnarray*}
4 \langle xy|\tilde{Z}_{EW}|xy \rangle \langle x|x \rangle  & = & 
 3 \, | x_2 x_0 y_1^* - x_1 x_2 y_0^*|^2  
 + 3 \, | x_1 x_1^* y_0 - x_2 x_0^* y_2|^2  
 + 3 \, | x_2 x_2^* y_2 - x_0 x_2^* y_0|^2  
 + 3 \, | x_2 x_1^* y_2 - x_1 x_1^* y_1|^2
   \\* 
& &  +      | 2 x_0 x_0^* y_2 - 2 x_1 x_2^* y_1 + x_2 x_2^* y_2 - x_0 x_2^* y_0|^2  
 +      | 2 x_0 x_1^* y_0 - 2 x_2 x_2^* y_1 + x_2 x_1^* y_2 - x_1 x_1^* y_1|^2  
 \\*
& &  + |2 \, x_0 x_1 y_2^* - x_2 x_0 y_1^* - x_1 x_2 y_0^*|^2  
 +      | 2 x_0 x_0^* y_0 - 2 x_1 x_0^* y_1 + x_1 x_1^* y_0 - x_2 x_0^* y_2|^2  \geq  0. 
\end{eqnarray*}
\end{widetext}
The expected value on the original state is $ \mbox{Tr}[\tilde{Z}_{EW}
\rho_\alpha] = \frac{1}{7} (3-\alpha)$, demonstrating entanglement for
all $\alpha > 3$.  Applying the non-PPT tests to this state fails to
show entanglement for $\alpha \alt 3.84$, even if we apply the
$6^{th}$ test, showing that this hierarchy can be considerably weaker
than the PPT hierarchy.

\subsection{$4 \otimes 4$ state}

We consider next the $4 \otimes 4$ state given by~\cite{footgisin}:
\[
\rho_\alpha = \frac{1}{2+\alpha} (|\psi_1 \rangle \langle \psi_1| +
|\psi_2 \rangle \langle \psi_2| + \alpha \cdot \sigma), \quad \alpha
\geq 0,
\]
where
\begin{eqnarray*}
\psi_1 &=& \frac{1}{{2}} (
|0 0 \rangle +
|1 1 \rangle +
\sqrt{2} \, |2 2 \rangle ), \\
\psi_2 &=& \frac{1}{{2}} (
|0 1 \rangle +
|1 0 \rangle +
\sqrt{2} \,  |3 3 \rangle ), \\
\sigma &= &
\frac{1}{8}(
|0 2 \rangle \langle 0 2 | +
|0 3 \rangle \langle 0 3 | +
|1 2 \rangle \langle 1 2 | +
|1 3 \rangle \langle 1 3 | +  \\
& & +
|2 0 \rangle \langle 2 0 | +
|2 1 \rangle \langle 2 1 | +
|3 0 \rangle \langle 3 0 | +
|3 1 \rangle \langle 3 1 |).
\end{eqnarray*}
Applying the PPT criterion yields provable entanglement only for those
states with $\alpha < 2 \sqrt{2} \approx 2.82843$. It was
suspected~\cite{footgisin} that the state was actually entangled for
all nonnegative values of $\alpha$. Using our new criteria, we show
that this is indeed the case, and provide an explicit entanglement
witness and its decomposition. Again, only the second level of our
hierarchy is needed. Using essentially the same approach as in the
example above, from the dual solution of the semidefinite program we
identify a particular witness:
\begin{widetext}
\begin{eqnarray*}
W &=&
(|2 2  \rangle - |0 0 \rangle) (\langle 2 2| - \langle 0 0 |) +
(|2 2  \rangle - |1 1 \rangle) (\langle 2 2| - \langle 1 1 |)  +
(|3 3  \rangle - |0 1 \rangle) (\langle 3 3| - \langle 0 1 |) +
(|3 3  \rangle - |1 0 \rangle) (\langle 3 3| - \langle 1 0 |) + \\*
&& + |2 3 \rangle \langle 2 3 | +
|3 2 \rangle \langle 3 2 | -
|2 2 \rangle \langle 2 2 | -
|3 3 \rangle \langle 3 3 |.
\end{eqnarray*}
This witness is nonnegative on all product states, as the following
identity certifies:
\begin{eqnarray*}
\langle xy|W|xy \rangle \langle x|x \rangle  &=&
 |x_0 x_0^* y_0+x_1 x_1^* y_0-x_2 x_0^* y_2-x_3 x_1^* y_3|^2+ 
 |x_0 x_0^* y_1+x_1 x_1^* y_1-x_2 x_1^* y_2-x_3 x_0^* y_3|^2+ \\ &&
 |x_2 x_2^* y_2+x_3 x_3^* y_2-x_0 x_2^* y_0-x_1 x_2^* y_1|^2+ 
 |x_2 x_2^* y_3+x_3 x_3^* y_3-x_1 x_3^* y_0-x_0 x_3^* y_1|^2+ \\ &&
 |                 x_1 x_3 y_2^*-x_0 x_2 y_3^*|^2 +
 |                 x_0 x_3 y_2^*-x_1 x_2 y_3^*|^2 +           
 |                 x_1 x_2 y_0^*-x_0 x_2 y_1^*|^2 +
 |                 x_0 x_3 y_0^*-x_1 x_3 y_1^*|^2 \geq 0.
\end{eqnarray*}
\end{widetext}
Applying the witness $W$ to the state, we obtain $ \mbox{Tr}[W
\rho_\alpha] = - \frac{2 (\sqrt{2}-1)}{2 + \alpha} < 0 $, therefore
certifying entanglement for {\it all} values of $\alpha$ in the
allowable range.

We have applied the second test of the hierarchy to many bound
entangled states found in the literature of dimensions up to 6 by 6 .
In all cases, this test has been {\it sufficient} to demonstrate
entanglement and construct numerical (and in some cases analytical)
entanglement witnesses.  However, we know from complexity arguments
that there must be states that pass the second test in the hierarchy
but are nonetheless entangled as we will discuss in the next section.

\section{Properties of the sets of entangled states with PPT symmetric extensions}

For every $k$, let us consider the set of states that are not detected
by the $k^{th}$ test of the hierarchy, which can also be characterized
as the states having PPT symmetric extensions to $k$ copies of party
$A$. They generate a sequence of nested sets, each one containing the
set of separable states.  The completeness theorem tells us that this
sequence actually converges to the set of separable states. It is
natural then to try to understand the properties of these particular
sets.  First of all, it is not difficult to see that these sets are
all convex and compact. But there are other interesting questions we
can ask about them. Are they nonempty? Is this sequence infinite or
does it collapse to a finite number of steps for certain cases? What
is the volume of the subset of entangled states contained in each set?
Are these sets invariant under LOCC? In this section we will address
some of these questions by explicitly constructing states with PPT
symmetric extensions and studying the implications of their existence.

\subsection{Constructing entangled states with PPT symmetric extensions}

We will now show explicitly how to construct an entangled state that
passes the second test in the hierarchy, which means it has a PPT
symmetric extension to two copies of system $A$.  We will proceed by
studying the properties of an EW obtained from the second test under a
particular scaling and then use duality arguments to infer the
existence of the required entangled state. The procedure is based on a
scaling technique developed recently by Reznick~\cite{reznick2003a}.

Let $\rho$ be a PPT entangled state that is detected by the second
test of the hierarchy. Then we know that there is an entanglement
witness $Z$ that satisfies $\mathrm{Tr}[\rho Z] < 0$ and
\begin{equation}
\langle x y | Z | x y \rangle \langle x | x \rangle\ \mathrm{is \ a}\ SOS,
\end{equation}
which implies that $\mathrm{Tr}[Z \rho_{sep}] \geq 0$ for any
separable state $\rho_{sep}$. Since $\rho$ is PPT, we know that the
bihermitian form $\langle x y | Z | x y \rangle$ cannot be a SOS,
otherwise $Z$ can easily be shown to be decomposable and hence unable
to detect a PPT entangled state, which would be a contradicition. Now,
let us consider the following hermitian operator
\begin{equation}
\label{Zgamma}
Z_\gamma = (A^{-1})^{\dag}\otimes \openone_B\, Z\, A^{-1}\otimes \openone_B,
\end{equation}
with $A = diag(1,{\gamma},\ldots,{\gamma})$, $\gamma > 0$. This
operator satisfies
\begin{equation}
\label{Zgammapos}
\langle x y | Z_\gamma | x y \rangle = \langle (A^{-1} x) y |\,Z\,| (A^{-1} x) y \rangle
\geq 0,
\end{equation}
since $Z$ is positive on product states. Now consider the state
\begin{equation}
\label{rhogamma}
\rho_\gamma = \frac{1}{N} (A \otimes \openone_B )\,\rho\,(A^{\dag} \otimes \openone_B ),
\end{equation}
with $N$ a positive normalization constant. The state $\rho_\gamma$ is
entangled for all $\gamma > 0$, because
\begin{equation}
\mathrm{Tr}[\rho_\gamma Z_\gamma] = \frac{1}{N}\mathrm{Tr}[\rho Z] < 0.
\end{equation}
This, together with (\ref{Zgammapos}), proves that $Z_\gamma$ is
actually an entanglement witness. Now, let us assume that $Z_\gamma$
is $\mathit{1}$-$SOS$ for all $\gamma > 0$, that is
\begin{equation}
\label{Zgamsos}
\langle x y | Z_{\gamma} | x y \rangle \langle x | x \rangle\ \mathrm{is \ a}\ SOS.
\end{equation}
By using (\ref{Zgamma}) and introducing the variables ${\tilde x} =
A^{-1} x$, we obtain
\begin{equation}
\label{sosscaled1}
\langle {\tilde x} y |Z| {\tilde x} y \rangle \langle 
A {\tilde x} | A {\tilde x} \rangle \ \mathrm{is \ a} \ SOS,
\end{equation} 
where the SOS structure is preserved due to the linearity of the
transformation from $x$ to ${\tilde x}$. Specializing for $A$ in
(\ref{sosscaled1}) we get
\begin{equation}
\label{sosscaled2}
\langle {\tilde x} y |Z| {\tilde x} y \rangle (
({\tilde x}_1^2 + \gamma^2 ({\tilde x}_2^2 +\ldots +{\tilde x}_{d_A}^2))
\ \mathrm{is \ a} \ SOS.
\end{equation}   
Since the set of forms that can be written as a SOS is a closed set,
the assertion in (\ref{sosscaled2}) must be valid also for $\gamma =
0$, and in that case we must have that
\begin{equation}
\langle {\tilde x} y |Z| {\tilde x} y \rangle {\tilde x}_1^2 \  \mathrm{is \ a}\  SOS.
\end{equation} 
But this implies that the form $\langle {\tilde x} y |Z | {\tilde x} y
\rangle$ is itself a SOS, which contradicts our assumption. Then we
conclude that there must be a value $\gamma^{\ast} > 0$ such that for
all $\gamma < \gamma^{\ast}$, the entanglement witness $Z_\gamma$ has
the property that
\begin{equation}
\label{nosos}
\langle x y |Z_\gamma |x y\rangle \langle x|x\rangle \ \mathrm{is\ not \ a} \  SOS.
\end{equation}
Note that this argument is only based in the fact that there is a PPT
entangled state detected by the second test of the hierarchy. Whenever
this happens, we know that we can construct an EW that satisfies
(\ref{nosos}), regardless of the dimensionality of the subsystems.

We can now very easily show that the existence of an EW satisfying
(\ref{nosos}) implies the existence of an entangled state that is not
detected by the second test.  We just need a very simple lemma of
convex analysis.  \newtheorem{lemma}{Lemma}
\begin{lemma}
  Let $K_1$ and $K_2$ be two closed convex cones such that $K_1
  \subset K_2$, where $\subset$ represents strict inclusion.  Then the
  dual cones satisfy $K_2^{\ast} \subset K_1^{\ast}$.
\end{lemma}
\textbf{Proof:} Recall the definition of the dual of a cone $K$,
$K^{\ast}=\{ z | \langle z,x \rangle \geq 0, \forall x \in K \}$.  Let
$z \in K_2^{\ast}$, then $\langle z,x \rangle \geq 0, \forall x \in
K_2$.  Since $K_1 \subset K_2$, $\langle z,x \rangle \geq 0, \forall x
\in K_1$, so $z \in K_1^{\ast}$ and hence $K_2^{\ast} \subseteq
K_1^{\ast}$. Now, let ${\tilde x} \in K_2$ such that ${\tilde x}
\notin K_1$. Assume that $K_2^{\ast} = K_1^{\ast}$. Let $z \in
K_1^{\ast}$; then we also have $z \in K_2^{\ast}$ and so $\langle
z,{\tilde x}\rangle \geq 0$. But since this is true $\forall z \in
K_1^{\ast}$, this means that ${\tilde x} \in K_1^{\ast \ast} = K_1$,
since $K_1$ is closed, and this is a contradiction. Then we must have
$K_2^{\ast} \subset K_1^{\ast}$. $\Box$

In our case we have the closed convex cones $O_k = \{ Z\, :\, \langle
x y|Z|x y\rangle (\sum_i |x_i|^2 )^k \ \mathrm{is \ a}\, SOS\}$,
$k=0,1,\ldots$, that we have already defined for the proof of Theorem
2. In this section we have shown that if there is an entangled state
that is detected by an EW in $O_1$ that is not in $O_0$, then $O_1$ is
\emph{strictly contained} in $O_2$. According to the lemma above this
means that $O_2^{\ast}$, which is the set of states that \emph{are
  not} detected by the third test of the hierarchy, is strictly
contained in $O_1^{\ast}$, the set of states that \emph{are not}
detected by the second test.  Thus, that there has to be a state that
\emph{is not} detected by an EW in $O_1$, because it belongs to
$O_1^{\ast}$ (equivalently, it passes the second test), but \emph{is}
detected by an EW in $O_2$, (because it does not belong to
$O_2^{\ast}$) and hence is entangled.

The discussion above shows that there has to be an entangled state
that passes the second test, but it does not give an explicit
construction. But since we know that $Z_\gamma$ ceases to be in $O_1$
for some small value of $\gamma$, and since we have shown that it
detects the entanglement of $\rho_\gamma$ for all $\gamma > 0$, we
could be tempted to say that then $\rho_\gamma$ has to be a state that
passes the second test for small enough $\gamma$. The problem is that
even though $Z_\gamma$ is an EW for $\rho_\gamma$, it is not clear
that there is not another EW in $O_1$ that detects the entanglement of
$\rho_\gamma$.  However, even if we are not assured that it would be
the case, we can still check whether $\rho_\gamma$ actually passes the
second test for small $\gamma$.  We did this numerically using our
code for the case of the Choi state (\ref{choiform}) with $\alpha =
3.0001$, and we found that indeed there is a value
$\gamma^{\ast}\simeq 0.4901$ of the parameter such that for all
$\gamma < \gamma^{\ast}$, the state $\rho_\gamma$ is entangled but
cannot be detected by the second test of the hierarchy.

\subsection{Properties of the hierarchy}

>From an algebraic point of view, the result of the previous subsection
is related to the fact that the fixed multiplier approach to proving
nonnegativity by finding a SOS decomposition does not behave well
under linear transformations. A solution to this problem may be
allowing the multiplier $\langle x| x\rangle$ to vary as well,
although it appears that this approach to checking for entanglement
cannot be stated as a SDP.

>From a physical point of view, this result has a very interesting
interpretation. The transformation represented by Equation
(\ref{rhogamma}) corresponds to applying an element of a POVM that
acts locally on system $A$ and leaves system $B$ alone. Such a
transformation can be implemented by local operations with some finite
probability. We see then that by SLOCC (stochastic local operations
and classical communication), we can transform a state that is
detected by the second test into a state that is not. Moreover, since
the matrix $A$ in (\ref{rhogamma}) is invertible, the reverse
transformation is \emph{also possible under SLOCC}. Then we could
start with the state $\rho'$, whose entanglement is not detected by
the second test, and by LOCC operations obtain, with some probability,
a state $\rho$ that \emph{is} detected by the second test. This shows
clearly that, unlike the PPT class of states, the classes of states
derived from the second and higher tests of the hierarchy \emph{are
  not invariant under LOCC}.

This scaling behavior of both states and entanglement witnesses is
very general and has very important consequences on the hierarchy of
tests. First, note that if we assume that $Z_\gamma$ is $k$-$SOS$ (for
any fixed $k$) for all $\gamma > 0$, this will be equivalent to
replacing $\langle x|x\rangle$ by $\langle x|x\rangle^k$ in
(\ref{Zgamsos}). We can then follow the exact same steps discussed
after (\ref{Zgamsos}) and arrive to the same contradiction (i.e., that
$\langle {\tilde x} y |Z | {\tilde x} y \rangle$ is a SOS). Then for
$\gamma$ small enough, $\langle x y | Z_{\gamma} | x y \rangle \langle
x | x \rangle^k$ must cease to be a SOS for \emph{any} fixed value of
$k$.  By applying Lemma 1 again, we can conclude that \emph{if there
  exists a PPT-entangled state} $\rho$ \emph{in} ${\cal H}_A \otimes
{\cal H}_B$, \emph{then for any value of k, there must be an entangled
  state} $\rho_k$ \emph{that is not detected by the} $k^{th}$
\emph{test of the hierarchy}. Note that this result depends only on
the existence of at least one bound entangled state, and not on the
dimensions of the system. Since there are explicit examples of bound
entangled states in $3\otimes 3$ (like Equation (\ref{choiform})) and
$2\otimes 4$ (see~\cite{horodecki1997a}), and we can (by embedding)
use them to construct bound entangled states in $N\otimes M$, $N\times
M>6$, we can conclude that \emph{there are always entangled states in}
${\cal H}_A \otimes {\cal H}_B$, $d_A \times d_B > 6$, \emph{that pass
  the first k tests in the hierarchy, for any fixed k}. In other
words, the hierarchy \emph{never collapses to a finite number of
  steps, even for fixed dimensions} (except in the already known cases
of $2\otimes 2$ and $2\otimes 3$).

A very interesting question has to do with what is actually the volume
of the set of entangled states that are detected by the $k^{th}$ test,
but are not detected by the $(k-1)^{th}$ test (or equivalently, the
set of states that have PPT symmetric extensions to $(k-1)$ copies of
$A$, but not to $k$ copies). Even though we cannot give any estimate
on the value of this volume, we can assert that this volume is
\emph{finite}, i.e., the set in question has \emph{nonzero measure}
when we consider the measure on the set of states introduced
in~\cite{zyczkowski1998a}.  This is true for all values of $k$ for
which this set is nonempty. The proof of this fact is a
straightforward translation of the result presented in Lemma 7
of~\cite{zyczkowski1998a}. This lemma proves that if a convex set
$C_1$ strictly contains a compact convex set $C_2$ that itself
contains a nonempty ball, then $C_1$ contains a nonempty ball that
does not intersect $C_2$.  In our case we can take $C_1$ to be the set
of states with PPT symmetric extensions to $(k-1)$ but not to $k$
copies of $A$, and $C_2$ the ones with extensions to $k$ copies.
Since both $C_1$ and $C_2$ are convex and compact, and $C_2$ contains
the set of separable states that contains a nonempty ball, the lemma
proves that there is a nonempty ball of states that have PPT symmetric
extensions to $(k-1)$ copies of $A$, but not to $k$ copies.

\section{Constructing bound entangled states from indecomposable
  entanglement witnesses}

In previous sections we have discussed how to use semidefinite
programs to implement separability criteria, and in particular we
showed how to exploit the duality of the SDP to generate an
indecomposable entanglement witness from a bound entangled state. In
this section we will show that we can also use a SDP to test whether a
given entanglement witness is decomposable. If the EW is
indecomposable, the dual program constructs a bound entangled state
that is detected by the witness.  The results of this section were
reported in~\cite{doherty2003b} which we follow closely.

As discussed above, a sufficient but not necessary condition for any
hermitian matrix $Z$ to be an entanglement witness is for it not to be
positive but rather decomposable as $Z = P+Q^{T_{A}}$, where $P \geq
0, Q \geq 0$. Such entanglement witnesses are obtained whenever a
state fails the first test of the hierarchy, which is just the PPT
criterion. These entanglement witnesses can only detect entangled
states that have a non-positive partial transpose. As it was shown in
(\ref{decompform}), the bihermitian forms associated with them can be
written as a SOS.

If we know only the matrix elements of $Z$ it may not be clear how to
determine whether $Z$ is decomposable or not. Consider the following
semidefinite program in the dual form
\begin{eqnarray}
\label{sdpindec}
{\rm maximize} &\ \  -{\mathrm{Tr}}\,[P+Q^{T_A}]/d_A d_B  \nonumber \\
{\rm subject\ to} &\ \ \ \ P \geq 0, \quad Q \geq 0 \nonumber \\
&\ \ H(P+Q^{T_A}) = H(Z),
\end{eqnarray}
where $H(Y)=Y-({\rm Tr}[Y])\openone/d_Ad_B$ is a linear map that
outputs the traceless part of $Y$. We can make (\ref{sdpindec}) take
the more familiar form (\ref{dual}) if we introduce the matrix
variable $X$, defined by $X=P\oplus Q$ ($X$ will play the role of $Z$
in (\ref{dual})). The final equality constraint may be enforced by a
finite number ($d_A^2d_B^2-1$) of trace constraints that define the
matrices $F_i$ and the coefficients $c_i$. $F_0$ is then proportional
to the identity.  By adding a sufficiently large multiple of the
identity to any matrix satisfying the trace constraints, it is always
possible to construct an $X = P\oplus Q >0$ also satisfying the
constraints. This means that the optimization is strictly feasible.
Let $\eta$ be the optimum value of the objective function $-{\rm
  Tr}\,[P+Q^{T_A}]/d_A d_B$, and $P_{\rm opt}$, $Q_{\rm opt}$ be the
values of $P$ and $Q$ that achieve this optimum. If
\begin{equation}
\label{etacond}
\eta \geq -{\mathrm{Tr}} \left[Z/d_A d_B \right],
\end{equation}
then for $\epsilon\equiv {\rm Tr} [Z/d_Ad_B] + \eta \geq 0$ it is
clear that we can write
\begin{equation}
\label{Zdecomp}
Z=(P_{\mathrm{opt}}+\epsilon \openone_{AB})+Q_{\mathrm{opt}}^{T_A},
\end{equation}
which shows that $Z$ is decomposable.

We have stated the semidefinite program in its dual form. The primal
form is worth considering since in the case where $Z$ is
nondecomposable, it constructs bound entangled states that are
detected by $Z$. Using the formulae (\ref{sdp}),(\ref{dual}) and
(\ref{sdpindec}), the primal form may be shown to be
\begin{eqnarray}
\label{primalindec}
\mathrm{minimize} &\ \  \mathrm{Tr}\,[Z\rho] - \mathrm{Tr}\,[Z/d_Ad_B] 
\nonumber \\
\mathrm{subject\ to} &\ \ \ \ \rho \geq 0, \quad \rho^{T_A} \geq 0 \nonumber \\
&\ \ \mathrm{Tr}\,[\rho] = 1,
\end{eqnarray}
where now the variables are the components of the state $\rho$ in some
basis.  The completely mixed state is strictly positive and is a
feasible solution of (\ref{primalindec}). Thus, because of the strict
feasibility of both the primal and dual problems the optima of these
two programs are equal~\cite{VaB:96} and there are matrices $\rho_{\rm
  opt},P_{\rm opt}, Q_{\rm opt}$ achieving the optimum. By
complementary slackness the range of $P_{\rm opt}$ is orthogonal to
the range of $\rho_{\rm opt}$ and the range of $Q_{\rm opt}$ is
orthogonal to the range of $\rho_{\rm opt}^{T_A}$.

Suppose now that the optimum $\eta$ satisfies $\eta < -{\rm Tr}
[Z/d_Ad_B] $. This, together with $\eta ={\rm Tr} [Z\rho_{\rm
  opt}]-{\rm Tr} [Z/d_Ad_B]$, means that
\begin{equation}
\label{trZrhoopt}
\mathrm{Tr}\,[Z\rho_{\rm opt}] <0.
\end{equation}
Since we know that $Z$ is an EW, then (\ref{trZrhoopt}) means that the
state $\rho_{\rm opt}$ is entangled. Furthermore, since this state is
a feasible solution of (\ref{primalindec}), it must satisfy $\rho_{\rm
  opt}^{T_A}\geq 0$, so this state is bound entangled. For any $P\geq
0$, $Q\geq 0$ we have
\begin{equation}
\label{Znondec}
{\mathrm{Tr}}\,[(P+Q^{T_A}) \rho_{\rm opt}] = {\mathrm{Tr}}\,[P \rho_{\rm opt}] 
+{\mathrm{Tr}}\,[Q \rho_{\rm opt}^{T_A}] \geq 0,
\end{equation}
so $Z$ cannot be decomposable, since it satisfies (\ref{trZrhoopt}).

We will now show that $\rho_{\rm opt}$ is the so-called edge PPT
entangled state. Since $\rho_{\rm opt}$ is a PPT entangled state, we
can write $\rho_{\rm opt}=(1-p)\rho_{sep}+p\delta$, where $\rho_{sep}$
is separable, $\delta$ is a so-called edge PPT entangled state and
$p>0$ is the minimum value for which such a decomposition is
possible~\cite{lewenstein2001a}. An edge PPT entangled state $\delta$
has the property that for any pure product state $|x,y\rangle$ and
$\epsilon >0$, $\delta-\epsilon |x,y\rangle\langle x,y|$ is either not
positive or not PPT. Since ${\rm Tr}\,[Z\rho_{sep}]\geq 0$ (because
$Z$ is an EW), if $p<1$ then
\begin{equation}
\label{contropt}
{\mathrm{Tr}}\,[Z\rho_{\rm
  opt}] > {\mathrm{Tr}}\, [ Z\delta ].
\end{equation}
But (\ref{contropt}) contradicts the optimality of $\rho_{\rm opt}$,
unless $\rho_{\rm opt}$ is itself an edge PPT entangled state.

This SDP finds the canonical decomposition of an indecomposable EW
discussed in~\cite{lewenstein2001a}.  Defining $\epsilon=-{\rm
  Tr}[Z\rho_{\rm opt}]>0$, we have
\begin{equation}
\label{Znondec2}
 Z=P_{\rm opt}+Q^{T_A}_{\rm opt}-\epsilon \openone_{AB},
\end{equation}
and as a result of the original dual form of the optimization,
$\epsilon$ is the smallest value for which such an expression holds
with $P\geq 0$, $Q\geq 0$. The range properties of $P_{\rm opt},Q_{\rm
  opt}$ and $\rho_{\rm opt}$ mean that this is the canonical form for
$Z$ introduced by Lewenstein {\em et al.}\cite{lewenstein2001a}.

\section{Characterization of positive maps}

It has been known for quite some time that there is a close
relationship between entanglement witnesses, positive bihermitian
forms and positive maps~\cite{choi1975a,woronowicz1976a}. In
particular, this relationship was exploited in~\cite{horodecki1996a}
to give a complete characterization of the separability problem in
terms of positive maps. We will now show how to translate the
properties of the entanglement witnesses generated by our hierarchy of
separability tests into a characterization of the set of strictly
positive maps.

Let us denote by ${\cal A}_A$ and ${\cal A}_B$ the set of linear
operators acting on ${\cal H}_A$ and ${\cal H}_B$ respectively. We
will call ${\cal L} ({\cal A}_A,{\cal A}_B)$, the set of linear maps
from ${\cal A}_A$ to ${\cal A}_B$. We say that a map $\Lambda \in
{\cal L} ({\cal A}_A,{\cal A}_B)$ is {\it positive}, if for any
operator $L \in {\cal A}_A$, $L \geq 0$, then $\Lambda (L) \geq 0$. A
{\it completely positive} (CP) map, is a map $\Lambda$ such that the
induced map
\begin{equation}
\label{cpmap}
\Lambda_n = \Lambda \otimes {\openone}_n : {\cal A}_A \otimes {\cal M}_n 
\rightarrow {\cal A}_B \otimes {\cal M}_n ,
\end{equation}
is positive for all $n$, with ${\cal M}_n $ being the space of
operators in a Hilbert space of dimension $n$ and ${\openone}_n$ the
identity map in that space.  CP maps have very important applications
in characterizing the set of physically meaningful evolutions of a
quantum state.

It is clear that any CP map is also a positive map. However, there are
positive maps that are not CP. This has very important consequences on
the study of entanglement of quantum states. In particular, there is a
one to one correspondence~\cite{horodecki1996a} between entanglement
witnesses and positive non-CP maps. Since the hierarchy of
separability tests offers a characterization of the interior of the
set of entanglement witnesses, it is not difficult to translate this
characterization to the set of positive non-CP maps. To do this, we
use the fact that for any linear operator $L \in {\cal A}_A \otimes
{\cal A}_B$, we can define a map $\Lambda \in {\cal L} ({\cal
  A}_A,{\cal A}_B)$ by
\begin{equation}
\label{posmap}
\langle k | \Lambda(|i\rangle \langle j|)|l\rangle =
\langle i|\otimes \langle k| L |j\rangle \otimes |l\rangle.
\end{equation}
Conversely, equation (\ref{posmap}) can be used to uniquely construct
the operator $L$ from the map $\Lambda$. Equivalently, we can
write~\cite{jamiolkowski1972a}
\begin{equation}
\label{posmap2}
\Lambda (\rho) = \mathrm{Tr}_A \left[ L (\rho^T  \otimes\openone_B ) \right],
\end{equation}
were $\rho$ is an operator in ${\cal A}_A$. Note that the same
operator $L \in {\cal A}_A \otimes {\cal A}_B$ can be used to define
two different maps in $ {\cal L} ({\cal A}_A,{\cal A}_B)$ and in $
{\cal L} ({\cal A}_B,{\cal A}_A)$.  It was
shown~\cite{jamiolkowski1972a} that this relationship gives in fact a
one to one correspondence between entanglement witnesses, i.e.,
hermitian operators that are positive on separable states but have a
negative eigenvalue, and positive non-CP maps.  By using
(\ref{posmap2}) it is not dificult to see that the interior of the set
of entanglement witnesses, which correspond to those $Z$ that satisfy
$\mathrm{Tr}[Z \rho_{sep}]>0$ for any separable state $\rho_{sep}$, is
mapped onto the set of positive maps that map any nonzero positive
semidefinite operator into a {\it positive definite} operator. Our
characterization of entanglement witnesses will translate into a
characterization of this subset of positive maps. The maps that are
left out are those that send at least one PSD operator into a another
PSD operator that is not positive definite.

In Section VI we showed that any $Z$ in the interior of the cone of
all entanglement witnesses is $k$-$SOS$ for some $k$. Since they
correspond to strictly positive maps (the ones that map any nonzero
PSD operator into a positive definite operator), we can characterize
these maps by associating a bihermitian form directly to the map,
using equation (\ref{posmap}).  Then we can state that a map is
strictly positive only if the form
\begin{eqnarray}
\label{mapform}
E_\Lambda(x,y) & = & \langle y|\Lambda (|x^* \rangle \langle x^*|) |y\rangle
\nonumber \\
& = &  \sum_{ijkl} (\langle k | \Lambda(|i\rangle \langle j|)|l\rangle)
 x^*_i y^*_k x_j y_l,
\end{eqnarray}
is $k$-$SOS$ for some value of $k$.

We can also give an interpretation of this characterization in a
language that only involves statements about maps. To do this we need
to analyze in more detail some of the properties of the EW generated
by the SDP. Let us consider the family of separability criteria that
searches for symmetric extensions of a certain state, but does not
require positive partial transposes. It is not difficult to see that
the entanglement witnesses generated by the second test will satisfy
\begin{equation}
\label{forms3}
\langle x y x|(Z_{EW} \otimes  \openone_A)|x y x\rangle =  \langle x y x|Z_0|x y
x\rangle,
\end{equation}
for all states $|x\rangle$ and $|y\rangle$, with $Z_0 \geq 0$. This is
the analogue to equation (\ref{forms2}). It is not difficult to show
that this equality implies that the operators $Z_{EW} \otimes
\openone_A$ and $Z_0$ actually {\it coincide} when they are restricted
to the symmetric subspace of the copies of system $A$. Furthermore,
this is true for any number of copies of system $A$. If we denote by
$\pi_k$ the projector onto the symmetric subspace of ${\cal
  H}_A^{\otimes k}$ (which we will denote by ${\cal H}_A^{\vee k}$),
we have
\begin{equation}
\label{ident}
\pi_k \otimes \openone_B (Z_{EW} \otimes  \openone_{A^{\otimes (k-1)}}) \pi_k 
\otimes \openone_B= (\pi_k\otimes \openone_B) Z_0 (\pi_k\otimes \openone_B).
\end{equation}
Since $Z_0$ is PSD on the space ${\cal H}_A \otimes {\cal H}_B \otimes
{\cal H}_A^{\otimes (k-1)}$, its restriction to the tensor product of
${\cal H}_A^{\vee k}$ and ${\cal H}_B$ remains PSD, which is the
right-hand side of (\ref{ident}). The completeness theorem of Section
V then tells us that if $Z_{EW}$ is a strictly positive entanglement
witness, then there must exist a finite $k$ for which equation
(\ref{ident}) is true.

We can now use the isomorphism defined by (\ref{posmap}) to restate
(\ref{ident}) in terms of properties of maps. First we use the fact
that this isomorphism gives a one to one correspondence between PSD
operators $L \in {\cal A}_A \otimes {\cal A}_B$ and CP maps $\Lambda
\in {\cal L} ({\cal A}_B, {\cal A}_A)$. Let $\Lambda:{\cal
  A}_B\rightarrow {\cal A}_A$ be the positive non-CP map associated
with $Z_{EW}$, and let $\bar\Lambda_k:{\cal A}_A \rightarrow {\cal
  A}_{{\cal H}_A^{\vee k}}$ be defined by $\bar\Lambda_k (\rho) =
\pi_k (\rho\otimes \openone_{A^{\otimes (k-1)}})\pi_k$. Equation
(\ref{posmap2}) can be used to check that the map associated with the
operator $\pi_k\otimes \openone_B (Z_{EW} \otimes \openone_{A^{\otimes
    (k-1)}}) \pi_k\otimes \openone_B$ is given by
\begin{equation}
\label{compmap}
\left( \bar\Lambda_k \circ \Lambda \right):{\cal A}_B\rightarrow
{\cal A}_{{\cal H}_A^{\vee k}}.
\end{equation}
But since the right-hand side of (\ref{ident}) is PSD, this map has to
be {\it completely positive}.

On the other hand, if $\Lambda$ is not a positive map, then the map
$\left( \bar\Lambda_k \circ \Lambda \right)$ cannot be completely
positive for any $k$. This is true because the map $ \bar\Lambda_k$
always maps a non-PSD matrix into a non-PSD matrix, as we can easily
show. Let $|i\rangle$ be an eigenvector of a non-PSD operator $\sigma$
in ${\cal H}_A$, with negative eigenvalue.  Then $\langle
i|\sigma|i\rangle <0$. For any $k$ the vector $|i\rangle^{\otimes k}$
belongs to the symmetric subspace ${{\cal H}_A^{\vee k}}$ and
satisfies $\pi_k |i\rangle^{\otimes k} = |i\rangle^{\otimes k}$. Then
we have
\begin{eqnarray}
\label{neg}
\left(\langle i|^{\otimes k}\right)\bar\Lambda_k (\sigma) \left(| i\rangle^{\otimes k}
\right)
& = & \left(\langle i|^{\otimes k}\right) \pi_k (\sigma\otimes
\openone_A^{\otimes k-1})\pi_k\left(| i\rangle^{\otimes k}\right) \nonumber \\
& = & \left(\langle i|^{\otimes k}\right) (\sigma\otimes
\openone_A^{\otimes k-1})\left(| i\rangle^{\otimes k}\right) \nonumber \\
& = & \langle i|\sigma|i\rangle <0,
\end{eqnarray}
and so $\bar\Lambda_k (\sigma)$ cannot be PSD.  Thus, we have the
following result: \newtheorem{thm3}[thm1]{Theorem}
\begin{thm3}
\label{posmapthm}
If the map $\Lambda:{\cal A}_B\rightarrow {\cal A}_A$ is strictly
positive, then there is a finite $k$ such that the map $\left(
  \bar\Lambda_k \circ \Lambda \right) :{\cal A}_B\rightarrow{\cal
  A}_{{\cal H}_A^{\vee k}}$ is completely positive. If for some $k$
the map $\left( \bar\Lambda_k \circ \Lambda \right)$ is completely
positive, then $\Lambda$ is a positive map.
\end{thm3}

Since this characterization of positive maps does not require solving
a SDP, because we only need to check positivity of a matrix, it is
interesting to study how efficient this approach is in actually
proving positivity of a map. To answer this question we consider the
following example based on the case of the $3\otimes 3$ state
considered in Section \ref{choiex}.  Let the map $\Lambda_0$ be
defined as $\Lambda_0(\rho) = \frac{1}{n} \mathrm{Tr}[\rho]
\openone_3$, where $\openone_3$ stands for the identity map in ${\cal
  H}_3$. The map $\Lambda_0$ lies in the interior of the cone of
positive maps.  Consider now a convex combination of $\Lambda_{0}$ and
the positive map $\Lambda_{\tilde{Z}_{EW}}$ induced by the witness in
equation (\ref{choiwit}), i.e.,
\[
\Lambda_{\alpha} = (1-\alpha) \Lambda_{0} + \alpha
\Lambda_{\tilde{Z}_{EW}}, \quad 0 \leq \alpha \leq 1.
\]
The map $\Lambda_{\tilde{Z}_{EW}}$ is in the boundary of the cone of
positive maps.  We have normalized the maps so that
$\Lambda_\alpha(\openone_3) = \openone_3$.  Since for $\alpha=0$ we
have $\Lambda_\alpha = \Lambda_0$ and for $\alpha=1$ we have
$\Lambda_\alpha = \Lambda_{\tilde{Z}_{EW}}$, the maps $\Lambda_\alpha$
are contained in a line segment with endpoints near the center and in
the boundary of the cone of positive maps, respectively. This implies
that $\Lambda_\alpha$ is a strictly positive map for $\alpha < 1$.

A natural question in this case is to determine the ranges of $\alpha$
for which we can effectively recognize positivity by applying the
result of Theorem \ref{posmapthm}. For this, as explained, we have to
form the tensor product of the given map with $k-1$ copies of the
identity, project on the symmetric subspace, and check whether the
resulting matrix is positive semidefinite. The computation of the
optimal $\alpha$ can be done in this case by solving a simple
generalized eigenvalue problem.

We have solved this numerically, for values of $k$ up to 8 (this
involves matrices of size $ 3 \cdot {3+k-1 \choose k}$, i.e., $135
\times 135$).  The obtained extreme values are shown in
Table~\ref{tab:kvals}, where $k$ is the number of extensions. The
results are consistent with the expected behavior $\lim_{k \rightarrow
  \infty} \alpha_k = 1$.

\begin{table}
\begin{center}
\begin{tabular}{c|c}
$k$ & $\alpha$ \\ \hline
1   &  0.4 \\
2   &  0.58769 \\
3   &  0.68556 \\
4   &  0.72727 \\
5   & 0.77663 \\
6   & 0.80766 \\
7   &  0.823529\\
8   & 0.846137
\end{tabular}
\caption{Number of extensions and optimal value of $\alpha$.}
\label{tab:kvals}
\end{center}
\end{table}
Notice that the convergence appears to be relatively slow, of order
$1/k$; in contrast, the SDP tests presented earlier based on the PPT
hierarchy can get all the way to the boundary $\alpha=1$ in just one
step.

It is interesting to note that Jamio{\l}kowski also studied the
problem of checking positivity of maps~\cite{jamiolkowski1974a}.  His
approach was related to ours in the sense that he showed that checking
positivity of a given map was equivalent to the nonnegativity of a
certain associated real polynomial. He then applied a general
technique for checking positivity of polynomials. As discussed in the
introduction there are several such algebraic methods and they all
scale badly with the problem size. In our case the specific problem of
checking positivity of a linear map between matrix algebras has been
reduced to a series of tests of matrix positivity, but none of them
succeeds uniformly for all maps. However, it is still the case that
for many instances, the positivity of a given linear map can be determined
and certified efficiently.

\section{Conclusions and discussion}

In this paper we have discussed a new family of separability criteria
for bipartite mixed states. Each criteria consists in searching for an
extension of a given state in a bigger space formed by adding a number
of copies of one of the subsystems, and requiring this extension to be
symmetric under exchanges of the copies and to remain positive under
any partial transpose. A failure to find such an extension proves
entanglement of the state, since it can be explicitly shown that
separable states have the required extensions. If an extension is
found, the test is inconclusive. This family of tests can be arranged
in a hierarchical structure, with each test being at least as powerful
as all the previous ones, and with the first test corresponding to the
well-known Peres-Horodecki PPT criterion.

This hierarchy of tests has two main properties that make it useful
and appealing.  First, the hierarchy is complete: any entangled state
will fail one of the tests at some finite point in the sequence.
Second, each test can be cast as a semidefinite program, which can be
efficiently solved. Furthermore, by exploiting the dual structure of
semidefinite programs, whenever a state is proven to be entangled by
failing one of the tests, an entanglement witness for that state can
be explicitly constructed. This duality can also help us to interpret
the hierarchy as trying to prove entanglement of a state by searching
for entanglement witnesses with a particular algebraic property that
states that the bihermitian form associated with the entanglement
witness can be written as a sum of squares when multiplied by a fixed
sum of squares to a certain power. The completeness of the hierarchy
can then be used to show that this algebraic property characterizes
all the elements in the interior of the cone of entanglement
witnesses.
 
We analyzed the computational resources needed to implement these
tests. We found that for a fixed test in the hierarchy, they scale
polynomially in the dimensions of the state. When we keep the size of
the state fixed, the resources also scale polynomially with the number
of copies added, or equivalently, with the order of the test in the
hierarchy. This behavior is very interesting in light of recent
results on the worst case complexity of the separability problem. It
has been shown that checking separability of a state is an NP-hard
problem when we study the scaling with respect to the dimensions of
both parties, so computational resources to solve it cannot scale
polynomially in this general case. In our family of tests this
non-polynomial behavior is reflected in how high up the hierarchy we
need to go to detect all entangled states.  Even though each test is
efficiently implementable, there are states for which we need to go
arbitrarily high in the hierarchy to show that they are entangled.

The dual formulation of the hierarchy can also be understood as the
construction of a sequence of cones, each one containing the previous
ones, that approximate the dual of the cone of separable states (which
contains the entanglement witnesses) from the inside, giving a
complete characterization of its interior.

We can also interpret the primal formulation as the construction of a
sequence of nested cones that approximate the cone of separable states
from the outside.  It is worth comparing this point of view with the
results in~\cite{woederman2003a}, where a semidefinite program was
used to approximate the cone of separable states from the inside. This
result however, only applies when one of the subsystems has dimension
2, and gives a complete characterization of separability only in this
particular case, while our hierarchy works for arbitrary dimensions of
the subsystems.

The hierarchy of tests allows us to divide the set of entangled states
into different classes, according to whether they have PPT symmetric
extensions to $k$ copies of one of the parties or not. This generates
a nested sequence of subsets of entangled states. This sequence can be
shown to be infinite for all dimensions of the subsystems, except for
$2 \otimes 2$ and $2\otimes 3$ where it is well-known that the PPT
criterion is enough to characterize entanglement (in these two special
cases, the hierarchy collapses to the first step). Furthermore, if the
set of states with PPT symmetric extensions to $(k-1)$ copies of $A$
but not to $k$ copies is nonempty, then it can be shown to have
nonzero measure. These classes of states however, are not closed under
LOCC operations, since we can transform a state that is not detected
by the second test into a state that is, with finite probability and
by applying only local operations.

\section{Acknowledgements}

It is a pleasure to acknowledge stimulating conversations with Hideo
Mabuchi, John Doyle, John Preskill and Ben Schumacher. Thanks to
Patrick Hayden for suggesting an improvement in an earlier proof of
Theorem 1, and to Nicolas Gisin for providing us with the PPT
entangled state in Section VII. B.  ACD gratefully acknowledges
conversations with Barbara Terhal.  FMS thanks Oscar Bruno for many
clarifying discussions regarding the completeness theorem. PAP
acknowledges interesting conversations with Bruce Reznick.  This work
was supported by the National Science Foundation as part of the
Institute for Quantum Information under grant EIA-0086083, the Caltech
MURI Center for Quantum Networks (DAAD19-00-1-0374) and the Caltech
MURI Center for Uncertainty Management for Complex Systems.

\appendix
\section{Improved SDP for implementing the tests}

We will now introduce a slight modification of the SDP given in
(\ref{sdptest}), that has the advantage of performing better
numerically. With $F$ given by (\ref{G}), let us consider the
following SDP
\begin{eqnarray}
\label{sdptest2}
{\mathrm{minimize}} &\ \  t  \nonumber \\
{\mathrm{subject \ to}} &\ \  t \openone_{ABA} + F(\mathbf{x})  \geq 0, 
\end{eqnarray} 
where $\openone_{ABA}$ is the identity matrix on the space ${\cal H}_A
\otimes {\cal H}_B \otimes {\cal H}_A$. It is clear that we can always
choose $t$ such that the LMI on the second line of (\ref{sdptest2}) is
satisfied. If the minimum of $t$ is negative or zero, then there
exists a value of $\mathbf{x}$ such that $F(\mathbf{x}) \geq 0$, which
is equivalent to say that (\ref{sdptest}) is feasible. On the other
hand, if the minimum of $t$ is strictly positive, then we know that
$F(\mathbf{x})$ cannot be PSD. Thus we see that feasibility of
(\ref{sdptest}) is equivalent to whether the minimum of
(\ref{sdptest2}) is strictly positive or not.  So we can use
(\ref{sdptest2}) to detect entangled states. This approach has the
property that the SDP (\ref{sdptest2}) is \textit{always feasible}.
This property makes the SDP solver to behave better numerically
(because it uses an interior point algorithm). This is in fact the SDP
that our code is solving when applying the tests to a given quantum
state.

\section{Strong duality and SDP infeasibility}
\label{apb}

We want to obtain infeasibility witnesses for the SDP
\[
F_0 + \sum_{i=1}^m x_i F_i \geq 0.
\]
Clearly, \emph{if} we can find a $W \geq 0$ such that
\[
\mathrm{Tr}[F_i W]= 0, \quad \mathrm{Tr}[F_0 W] < 0,
\]
then the SDP is necessarily infeasible, as follows by the argument
given after equation (\ref{feascert}). Under what conditions does such
a $W$ exist?  As we mentioned earlier, we need some form of strong
duality to hold.

Consider the set $\mathcal{S}: = \range \mathcal{F} + K$, where $K$ is
the PSD cone, $\mathcal{F}:\R^m \rightarrow \Sy^n$ is the linear map
defined by $\mathcal{F}(x) := \sum_{i=1}^m x_i F_i$, and $A + B = \{ y
| y = a + b, a \in A, b\in B \}$. Feasibility of the SDP is equivalent
to $F_0 \in \mathcal{S}$. The set $\mathcal{S}$ is obviously convex.
Now, \emph{if} $\mathcal{S}$ is also closed, then we can apply the
separating hyperplane theorem, and conclude the existence of a $W$ as
above.

The difficulty, of course, is than in general the sum of two closed
sets may not be closed. In particular, in SDP things can go wrong. For
instance, for
\[
\left[\begin{array}{cc} x & 1 \\ 1 & 0
\end{array}\right]
\geq 0
\]
which is obviously infeasible, it is not hard to see that \emph{no
  witness} $W \geq 0$ as above can exist. This can be traced back to
the fact that $\mathcal{S}$ in this case is not closed.

So, what conditions can be required to guarantee that $\mathcal{S}$ be
closed? An often-used criterion is the so-called \emph{Slater
  condition}~\cite{rockafellar1970a}, which in our case is the
following:

\vspace{.4cm} If $\ker \mathcal{F}^* \cap \ri K^* \not = \emptyset$,
then $\range \mathcal{F} + K$ is closed.  \vspace{.4cm}

Here, $\mathcal{F}^*$ is the adjoint map of $\mathcal{F}$, $K^*$ is
the dual cone (equal to $K$, in this case), and $\ri$ denotes the
relative interior of a set.

In other words, to guarantee the existence of infeasibility witnesses
of the form we described (for any possible $F_0$), it is sufficient to
show a $Z > 0$, that satisfies $\mathrm{Tr}[F_i Z] =0$, for all $i =1,
\ldots, m$. Notice that this \emph{almost} looks like the certificate
$W$ we are after, except that $F_0$ does not appear in the expression
(otherwise, the condition would be useless). In general, checking
whether the Slater condition is satisfied in concrete problems is not
too difficult.

For our SDPs in (\ref{sdptest}) and (\ref{dual2}), it is immediate to
show that the criterion is indeed satisfied, as all the matrices $F_i$
are traceless, so we can just take $Z = \openone > 0$.

\bibliographystyle{prsty} \bibliography{Cavsep}

\end{document}